\documentclass[aps,showpacs]{revtex4}

\usepackage{graphicx}
\usepackage{amsmath}
\usepackage{amssymb}

\begin{document}

%\preprint{HEP/123-qed}

\title[Carlson]{Effective action approach and
Carlson-Goldman mode in $d$-wave superconductors}

\author{Sergei~G.~Sharapov} \thanks{On leave of absence from Bogolyubov
 Institute for Theoretical Physics, Kiev, Ukraine}%
 \email{Sergei.Sharapov@unine.ch} \author{Hans~Beck}%
 \email{Hans.Beck@unine.ch}
 \homepage{http://www.unine.ch/phys/theocond/}
 \affiliation{Institut de Physique,
 Universit\'e de Neuch\^atel, 2000 Neuch\^atel, Switzerland}

\date{8 November, 2001}% It is always \today, today, but you may specify any
% date with \date.

\begin{abstract}
We theoretically investigate the Carlson-Goldman (CG) mode in two-dimensional
clean $d$-wave superconductors using the effective ``phase only'' action formalism.
In conventional $s$-wave superconductors, it is known that the CG
mode is observed as a peak in the structure factor of the pair
susceptibility $S(\Omega, \mathbf{K})$ only just below the transition
temperature $T_c$ and only in dirty systems. On the other hand,
our analytical results support the statement by Y.~Ohashi and S.~Takada,
Phys.~Rev.~B {\bf 62}, 5971 (2000) that in $d$-wave superconductors the CG
mode can exist in clean systems down to the much lower temperatures,
$T \approx 0.1 T_{c}$. We also consider the manifestations of the CG  mode
in the density-density and current-current correlators and discuss the gauge
independence of the obtained results.

\end{abstract}


%\keywords{Suggested keywords}%Use showkeys class option if keyword
                              %display desired
\maketitle

\section{Introduction}

More than 25 years ago, an unusual propagating sound-like
($\Omega/K = v_{\rm CG}$) Carlson-Goldman (CG) mode in {\em charged} superconducting
systems with the velocity $v_{\rm CG} = 10^3 \, -\, 10^4 m/c$
was discovered \cite{Goldman.discovery} (see also \cite{Goldman.1981}).
It was a widespread opinion before the CG mode discovery that since in
the charged systems the sound-like Bogolyubov-Anderson (BA)
\cite{Bogolyubov,Anderson}
mode associated with {\em neutral} superconductors is converted to
the plasma mode due to the Anderson-Higgs mechanism, there is no
sound-like phase mode in charged systems.

A magnificent effort (see, for example, the reviews
\cite{Artemenko,book.1980,Galaiko,book.1986}) was made to understand the mechanism,
responsible for the appearance of the CG mode and its relation to other
phenomena of non-equilibrium superconductivity. While the majority theories of the CG
mode \cite{Artemenko,book.1980,Galaiko,book.1986}) were essentially based
on the kinetic equations which are usually derived using the quasi-classical
Green's functions, the paper by Kulik, Entin-Wohlman, Orbach \cite{Kulik.1981}
used a more conventional approach based on the Matsubara Green's functions
without kinetic equations. In the subsequent papers of Takada with coauthors
(see \cite{Takada.1988,Takada.1997} and references therein)
the approach of \cite{Kulik.1981} was further developed and very recently applied
for the case of $d$-wave superconductivity \cite{Takada.2000}. 
The collective oscillations in $d$-wave superconductors were also studied
using the kinetic equations for Green's functions \cite{Artemenko.1997}
(see also \cite{Artemenko.2001}).

Since the discovery of high-$T_c$ compounds, the $d_{x^2-y^2}$ superconductivity
has attracted much attention \cite{d-wave} and the claim of \cite{Takada.2000} that
the CG mode in clean $d$-wave superconductors may survive in a much wider
region of temperatures down to $0.2T_c$ appears to be very different from the
established properties of the CG mode in $s$-wave superconductors, so that
it is important to check it by an independent calculation.

On the other hand, the importance of phase fluctuations in high-temperature
superconductors (HTSC) (see, e.g. the review \cite{Loktev.review}), stimulated
interest in the derivation of the ``phase only'' effective actions from
the microscopic theory. It is important to emphasize that although there
is no commonly accepted theory of HTSC, it seems reasonable that one can use
a simple BCS-like approach to describe the properties of HTSC
below the critical temperature, $T_c$ even though such an approach fails
above $T_c$. Relying on this argument the ``phase only'' actions for
$d$-wave superconductors were recently derived in
\cite{Randeria.action,we.LD,Benfatto}. However, the only phase excitations
which are described by these actions are the BA mode in the {\em neutral}
superconductor \cite{we.LD} and the plasma mode \cite{Randeria.action,Benfatto}
which appears when the Coulomb interaction is taken into account.
This corresponds to the standard paradigm which does not yield the
existence of the CG mode. Thus the purpose of the present paper is to investigate
which ingredient is missing in the treatments of
\cite{Randeria.action,we.LD, Benfatto}, so that the CG mode does not appear
in these approaches and to establish a link between the results
of \cite{Takada.2000} and the ``phase only'' action formalism. This missing link
is established here and the CG mode is obtained within the effective action formalism.
Our main results can be summarized as follows.

\indent
1. We extend the ``phase only'' effective action formalism for {\em charged}
systems to incorporate the {\em density-current} coupling which was so far
considered only using other methods
\cite{Kulik.1981,Takada.1988,Takada.1997,Takada.2000}
(see also \cite{Varlamov.1986,Smith.1995,Smith.2000}, where
the effect of this coupling appears to be important for the description
of dirty superconductors). In neutral systems the density-current
correlator {\em does contribute} in so called {\em Landau terms} of the effective
action  \cite{Aitchison.2000,we.LD}, so that the correct expression for these terms
can only be obtained when this correlator is taken into account.

2. We show that when the density-current coupling is included it
becomes possible to obtain the CG mode using the ``phase only''
action. In particular, we derive an analytical expression for the pair
susceptibility structure factor and solve numerically the equation for the CG
mode velocity.

\indent
3. We show the gauge independent character of the equation for the collective
phase excitations one of the solutions of which is the CG mode.
Establishing a link between the pair susceptibility and this gauge independent
equation for the phase excitations we argue that the peaks in the
structure factor associated with the CG mode are independent of the choice
of the gauge. The gauge independence of the equation for the CG mode used in
the previous papers \cite{Takada.1988,Takada.1997, Takada.2000} is also shown
applying the identity derived recently in \cite{Smith.2000}.

\indent
4. We consider possible manifestations of the CG mode in the gauge
independent density-density and current-current response functions.

\indent
5. We derive analytical expressions for the density-density,
current-current and density-current polarization functions
for 2D clean $d$-wave superconductor at $T \ll T_{c}$.

The paper is organized as follows:
In Sec.~\ref{sec:model} we introduce our model.
In Sec.~\ref{sec:action}, we describe all details about the formalism used in the paper,
necessary for further understanding.
The various forms of the effective actions (for the phase and electric potential,
``phase only'' and ``electric potential only'') are expressed in terms of
density-density, current-current and density-current polarization functions
in Sec.~\ref{sec:Omega}. The general expressions for these polarization functions
are given in Appendix \ref{sec:A}, their derivation for $d$-wave
superconductors is considered in Appendix~\ref{sec:C} and the nodal approximation
employed for this derivation is briefly discussed in Appendix~\ref{sec:B}.
In Sec.~\ref{sec:Omega}
we also discuss in detail the difference between the present and other
\cite{Kulik.1981,Takada.1988,Takada.1997,Takada.2000,Varlamov.1986,Smith.1995,Smith.2000}
approaches. In Sec.~\ref{sec:standard} we briefly recall the properties
of the phase excitations in the absence of the density-current coupling and stress
some points like gauge independence of the equation for the collective phase
excitations and the properties of the gauge independent density-density and
current-current correlators which are particularly useful for better understanding of our
main results which are presented in Sec.~\ref{sec:main}. In particular, in
Sec.~\ref{sec:CG.equation} we derive the equation for the CG mode, give
its physical interpretation and discuss the gauge independence of the present
and previous approaches. In Sec.~\ref{sec:velocity} we present the results
for the velocity of the CG mode and Sec.~\ref{sec:structure.factor} is devoted
to the structure factor (the calculational detail for these sections are given
in Appendix~\ref{sec:D}).
We conclude in Sec.~\ref{sec:discussion} with a discussion
and summary of our results.

\section{Model Hamiltonian}
\label{sec:model}
Let us consider the following action (in our notations
the functional integral is expressed via $e^{S}$)
\begin{equation}
\label{S}
S = - \int_{0}^{\beta} d \tau \left[\sum_{\sigma} \int
d^2r \psi_{\sigma}^{\dagger}(\tau, {\bf r})
(\partial_{\tau} - i e A_{0}(\tau, {\bf r}))
\psi_{\sigma}(\tau, {\bf r}) + H(\tau) \right]\,,
\qquad \mathbf{r} = (x,y)\,,
\qquad \beta \equiv \frac{1}{T}\,,
\end{equation}
where the Hamiltonian $H(\tau)$ is
\begin{equation}
\label{Hamilton}
\begin{split}
H(\tau) & = \sum_{\sigma} \int d^2 r \psi_{\sigma}^{\dagger}(\tau, {\bf r})
[\varepsilon(- i \nabla - \frac{e}{c} {\bf A}(\tau, {\bf r})) - \mu]
\psi_{\sigma}(\tau, {\bf r})      \\
& - \frac{1}{2} \sum_{\sigma} \int d^2 r_1 \int d^2 r_2
\psi_{\sigma}^{\dagger}(\tau, {\bf r}_2) \psi_{\bar{\sigma}}^{\dagger}(\tau,
{\bf r}_1) V({\bf r}_1; {\bf r}_2) \psi_{\bar{\sigma}}(\tau, {\bf r}_1)
\psi_{\sigma}(\tau, {\bf r}_2)\\
& + \frac{1}{2}  \int d^2 r_1 \int d^2 r_2
(\sum_{\sigma} \psi_{\sigma}^{\dagger}(\tau, {\bf r}_1)
\psi_{\sigma}(\tau, {\bf r}_1) -n  ) V_{c}({\bf r}_1 - {\bf r}_2)
(\sum_{\sigma^{\prime}} \psi_{\sigma^{\prime}}^{\dagger}(\tau, {\bf r}_2)
\psi_{\sigma^{\prime}}(\tau, {\bf r}_2) -n  )
\,.
\end{split}
\end{equation}
Here  $\psi_{\sigma}(\tau, {\bf r})$ is a fermion field with the spin
$\sigma= \uparrow, \downarrow$, $\bar{\sigma} \equiv - \sigma$,
$\tau$ is the imaginary time and $V({\bf r}_1; {\bf r}_2)$
is an attractive short-range potential, $ V_{c}({\bf r}_1 - {\bf r}_2)$
is the long range Coulomb interaction, $n$ is the neutralizing
background charge density. Throughout the paper we call the superconducting system
{\em neutral} if the last term of Eq.~(\ref{Hamilton}) is omitted and {\em charged}
if this term is taken into account. Even in the latter case the whole superconductor
remains, of course, neutral due to the neutralizing ionic background.

We assume that the momentum representation
of  $V({\bf r}_1;  {\bf r}_2)$ contains attraction only in the
$d$-wave channel (see the discussion in \cite{Randeria.action}). The Fourier transform
of the Coulomb interaction $V_{c}({\bf q})$ depends on the detail
of the model. It can, for example, be taken
$V_{c}({\bf q}) = 4 \pi e^2/ q^2$ in 3D,  $V_{c}({\bf q}) = 2 \pi e^2 /q$
in 2D or more complicated expression (see e.g. \cite{Takada.2000,Randeria.action})
if the layered structure of HTSC is taken into account.
However, the detailed expression is not crucial for the CG mode, because the mode
appears when the Coulomb interaction is screened out by the quasiparticles.
(The form of the expression would be, of course, essential for the analysis
of the plasma mode \cite{Takada.1998,Randeria.action}.)
The form of dispersion law, $\varepsilon(\mathbf{k})$,
is also not essential because the
final results for the $d$-wave case will be formulated in terms of
the non-interacting Fermi velocity
$\mathbf{v}_F \equiv \partial \varepsilon(\mathbf{k})/\partial
\mathbf{k}|_{\mathbf{k} = \mathbf{k}_F}$
and the gap velocity
$\mathbf{v}_\Delta \equiv \partial \Delta(\mathbf{k})/\partial
\mathbf{k}|_{\mathbf{k} = \mathbf{k}_F} $, where
$\Delta(\mathbf{k})$ is the momentum dependent superconducting gap.
We will also use the parameter $\alpha_{D} \equiv v_{F}/v_{\Delta}$
which is called the anisotropy of the Dirac cone.
Throughout the paper $\hbar = k_{B} = 1$ units are chosen.
An external electromagnetic field $A = (A_0, \mathbf{A})$ was introduced in the action
Eq.~(\ref{S}) to calculate various correlation functions using
the functional derivatives with respect to this external source field.

\section{Derivation of the effective action and the structure factor}
\label{sec:action}

The derivation of the effective ``phase only'' action for
neutral (see e.g. \cite{Aitchison.2000,we.LD,Loktev.review}) and
charged \cite{Nagaosa.book,Capezzali,Randeria.action,Benfatto}
$s$- and $d$-wave superconducting systems is widely discussed in the literature, so we
briefly recap the main steps, including the functional integral
representation for the structure factor, and
making in Sec.~\ref{sec:Omega} a point on the appearance  of the term which
couples density and current.

The first step of the derivation is to use the appropriate
Hubbard-Stratonovich transformations to decouple four-fermion
interaction terms in the attractive and repulsive channels.
The attractive part of the interaction was recently considered in detail
in Sec.~II of \cite{we.LD} using the bilocal Hubbard-Stratonovich
fields  $\Phi(\tau,\mathbf{r}_1; \mathbf{r}_2)$ and
$\Phi^{\dag}(\tau,\mathbf{r}_1; \mathbf{r}_2)$ (see \cite{Kleinert}
for a review), so we show explicitly the corresponding
transformation only for the Coulomb interaction:
\begin{equation}
\begin{split}
& \exp\left[ - \frac{1}{2} \int d \tau \int d^2 r_1 \int d^2 r_2
(\psi_{\sigma}^{\dagger}(\tau, \mathbf{r}_1) \psi_{\sigma}(\tau, \mathbf{r}_1)
- n) V_c (\mathbf{r}_1, \mathbf{r}_2)
(\psi_{\sigma^{\prime}}^{\dagger}(\tau, \mathbf{r}_2) \psi_{\sigma^{\prime}}(\tau, \mathbf{r}_2)
- n) \right] =\\
& \int \mathcal{D} \varphi
\exp \left\{ - \int d \tau  \int d^2 r_1 \int d^2 r_2
\left[ \frac{1}{2} e \varphi (\tau, \mathbf{r}_1)
V_c^{-1} (\mathbf{r}_1, \mathbf{r}_2)  e \varphi (\tau, \mathbf{r}_2) -
i  e \varphi (\tau, \mathbf{r}_1)
(\psi_{\sigma}^{\dagger}(\tau, \mathbf{r}_1) \psi_{\sigma}(\tau, \mathbf{r}_1)
- n) \delta(\mathbf{r}_1 - \mathbf{r}_2)  \right] \right\}\,,
\end{split}
\end{equation}
where the Hubbard-Stratonovich field $\varphi$ has the meaning of the electric
potential. Thus, the partition function is
\begin{equation}
\label{statistical.sum}
\begin{split}
Z = & \int \mathcal{D} \Psi^{\dagger} \mathcal{D} \Psi \mathcal{D} \Phi^{\dag}
\mathcal{D} \Phi \mathcal{D} \varphi
\exp[S(\Psi^{\dagger}, \Psi, \Phi^{\dagger}, \Phi, \varphi, A)], \\
 S & (\Psi^{\dagger}, \Psi, \Phi^{\dagger}, \Phi, \varphi, A) =
\int_{0}^{\beta} d \tau \int d^2 r_1 \int d^2 r_2
\left\{-\frac{1}{V(\mathbf{r}_1; \mathbf{r}_2)}
|\Phi (\tau,\mathbf{r}_1; \mathbf{r}_2)|^2
- \frac{1}{2} e \varphi (\tau, \mathbf{r}_1)
V_c^{-1} (\mathbf{r}_1, \mathbf{r}_2)  e \varphi (\tau, \mathbf{r}_2)
\right.  \\
& + \Psi^{\dagger}(\tau,\mathbf{r}_1)
[-\partial_{\tau}+ i e \tau_3 A_0 + i e \tau_3 \varphi(\tau, \mathbf{r}_1)-
\tau_{3}\xi(-i \nabla - \tau_3 \frac{e}{c} \mathbf{A})]
\Psi(\tau,\mathbf{r}_2) \delta(\mathbf{r}_1 - \mathbf{r}_2) \\
&  \left. - i e \varphi(\tau, \mathbf{r}_1) n \delta(\mathbf{r}_1 - \mathbf{r}_2)
+ \Phi^{\dag}(\tau,\mathbf{r}_1; \mathbf{r}_2)
\Psi^{\dagger}(\tau,\mathbf{r}_1) \tau_{-}\Psi(\tau,\mathbf{r}_2) +
\Psi^{\dagger}(\tau,\mathbf{r}_1) \tau_{+}\Psi(\tau,\mathbf{r}_2)
\Phi(\tau,\mathbf{r}_1; \mathbf{r}_2) \right\}
\end{split}
\end{equation}
where $\Psi$ and $\Psi^{\dagger}$ are the Nambu spinors,
$\xi(-i \tau_3 \nabla) \equiv \varepsilon(-i \tau_3 \nabla) - \mu$
and $\tau_3$, $\tau_{\pm} = (\tau_1 \pm i \tau_{2})/2$ are Pauli matrices.

To consider the Hubbard-Stratonovich field $\Phi$ it is convenient
to use the relative $\mathbf{r} = \mathbf{r}_1 - \mathbf{r}_2$
and center of mass coordinates $\mathbf{R} = (\mathbf{r}_1 + \mathbf{r}_2)/2$,
so that $\Phi(\tau, \mathbf{r}_1, \mathbf{r}_2) \equiv
\Phi(\tau, \mathbf{R}, \mathbf{r})$.
Now, using the functional integral representation, the imaginary time
pair susceptibility is defined as
\begin{equation}
\label{pair.susceptibility}
\eta(\tau, \mathbf{R}_1 - \mathbf{R}_2) = - \frac{1}{Z(A =0)}
\int \mathcal{D} \Psi^{\dagger} \mathcal{D} \Psi \mathcal{D} \Phi^{\dag}
\mathcal{D} \Phi \mathcal{D} \varphi
\Phi(\tau, \mathbf{R}_1, \mathbf{0}) \Phi^{\dagger}(0, \mathbf{R}_2, \mathbf{0})
\exp[S(\Psi^{\dagger}, \Psi, \Phi^{\dagger}, \Phi, \varphi)] \,.
\end{equation}
Since the distance $|\mathbf{R}_1 - \mathbf{R}_2|$ is expected to be larger
than the internal Cooper pair scale, it is possible to put
$\mathbf{r}=0$ in Eq.~(\ref{pair.susceptibility}).
The structure factor $S(\Omega, \mathbf{K})$ which used to present
the experimental data \cite{Goldman.1981,book.1980} is related to the real
frequency pair susceptibility $\eta(i \Omega_n \to \Omega + i0, \mathbf{K})$
by
\begin{equation}
\label{structure.factor}
S(\Omega, \mathbf{K}) = -2 [1- \exp(-\beta \Omega)]^{-1}
\mbox{Im} \eta(\Omega+i0, \mathbf{K}) \approx - \frac{2T}{\Omega}
\mbox{Im} \eta(\Omega+i0, \mathbf{K})\,.
\end{equation}
The definition of the pair susceptibility Eq.~(\ref{pair.susceptibility})
is apparently gauge dependent, since the auxiliary Hubbard-Stratonovich
field is gauge dependent. Nevertheless, as we discuss later the poles of
$\eta(\Omega, \mathbf{K})$ are {\em gauge independent} and this justifies
the use of Eqs.~(\ref{pair.susceptibility}) and (\ref{structure.factor})
to extract the observable values.

The simplest way to study the low energy phase dynamics
\cite{foot1} is to employ the canonical
gauge transformation \cite{foot2}
\begin{equation}
\label{gauge}
\Psi(x) \to
\begin{pmatrix}
e^{i \theta (x)/2} & 0 \\
o & e^{-i \theta (x)/2}
\end{pmatrix}
\Upsilon (x), \qquad x \equiv  (\tau, \mathbf{r})
\end{equation}
separating the phase of the ordering field
\begin{equation}\label{Phi.modulus.approx}
\Phi(\tau, \mathbf{R}, \mathbf{r}) \approx \Delta(\tau, \mathbf{R}, \mathbf{r})
\exp[i \theta(\mathbf{\tau, R})]\,.
\end{equation}
Then after the integration over the Fermi-fields the partition function
becomes
\begin{equation}
\label{statistical.sum.phase} Z = \int \Delta \mathcal{D} \Delta
\mathcal{D} \theta \mathcal{D} \varphi
\exp[-\beta \Omega\{ \Delta, \partial \theta, \varphi, A \}]\,,
\end{equation}
where the effective potential
\begin{equation}
\label{Omega}
\begin{split}
\beta & \Omega\{ \Delta, \partial \theta, \varphi, A \} = \\
& \int_{0}^{\beta} d \tau \int d^2 r_1 \int d^2 r_2 \left[
\frac{\Delta^{2}(\tau,\mathbf{R},\mathbf{r})}{V(\mathbf{r}_1 - \mathbf{r}_2)}
+ \frac{1}{2} e \varphi (\tau, \mathbf{r}_1)
V_c^{-1} (\mathbf{r}_1, \mathbf{r}_2)  e \varphi (\tau, \mathbf{r}_2)
+ i e \varphi(\tau, \mathbf{r}_1) n \delta(\mathbf{r}_1 - \mathbf{r}_2)
\right]
- \mbox{TrLn} G^{-1}
\end{split}
\end{equation}
with the inverse Green's function
\begin{equation}
\label{Green}
G^{-1} = \mathcal{G}^{-1} - \Sigma\,,
\end{equation}
\begin{equation}
\label{Sigma}
\begin{split}
& \langle \tau_1, \mathbf{r}_1|\Sigma (\partial \theta, \varphi, A)
| \tau_2, \mathbf{r}_2 \rangle = \left[
\tau_{3} \left(i \frac{\partial_{\tau_{1}} \theta}{2}
- i e \varphi(\tau_1, \mathbf{r}_1) - i e A_0 (\tau_1, \mathbf{r}_1) \right)
+ \tau_3 O_1 (\mathbf{r}_1) + \hat{I} O_2 (\mathbf{r}_1) \right]
\delta(\tau_1 - \tau_2) \delta(\mathbf{r}_1 - \mathbf{r}_2)\,.
\end{split}
\end{equation}
For $T \ll T_c$ it is reasonable to neglect the amplitude fluctuations
and assume that the amplitude of the order parameter
$\Delta(\tau, \mathbf{R}, \mathbf{r})$ does not depend on $\mathbf{R}$.
Then the frequency-momentum representation of  $\mathcal{G}$ in Eq.~(\ref{Green})
is the usual Nambu-Gor'kov Green's function
\begin{equation}\label{Green.neutral.momentum}
\mathcal{ G}(i \omega_{n},  {\bf k}) = - \frac{ i \omega_{n} \hat{I} +
\tau_{3} \xi( {\bf k}) - \tau_{1} \Delta(\mathbf{k})} {\omega_{n}^{2} + \xi^{2}(
{\bf k}) + \Delta^{2}(\mathbf{k})}\,,
\end{equation}
where, because $d$-wave pairing is considered
$\Delta(\mathbf{k}) = \Delta/2 (\cos k_x a - \cos k_y a)$
($a$ is the lattice constant)
and $\omega_n = \pi (2n+1)T$ is fermionic (odd) Matsubara frequency.
Since in what follows only the low temperatures, $T \ll \Delta(T)$
are considered, we can replace the temperature dependent amplitude $\Delta(T)$
by its zero temperature value, $\Delta_{0} \equiv \Delta(T =0)$.
Thus all linear low temperature dependences of the polarization functions
considered below are due to the nodes of $\Delta(\mathbf{k})$, but not to
the temperature dependence of $\Delta(T)$ itself.

The precise form of the operators $O_1$ and $O_2$ in Eq.~(\ref{Sigma})
which depends on the particular form of the tight-binding spectrum
$\varepsilon(\mathbf{k})$ is given in \cite{we.LD}
(see also \cite{Benfatto} for the formulation of the general rules for
representation of $\Sigma$). It is essential, however, that the coordinate
representation of $\Sigma$ does not depend on the phase $\theta$ itself
and contains only its derivatives. Thus the coordinate representation
of $\Omega_{kin}$ is also expressed via the derivatives of $\theta$.
This property is particularly convenient for studying 2D models when a constant
space independent phase is prohibited by the Coleman-Mermin-Wagner-Hohenberg (CMWH)
theorem.

Since we are interested only in the phase dynamics in the presence
of Coulomb interaction, in what follows
we consider only the phase $\theta$ and the electric potential
$\varphi$ dependent parts of the thermodynamical potential Eq.~(\ref{Omega}).
This part of $\Omega$ which we denote as $\Omega_{kin}$
can be present as a series
\begin{equation} \label{Omega.Kinetic.phase}
\begin{split}
\Omega_{kin} \{\partial \theta, \varphi \}
 =
\int_{0}^{\beta} d \tau \int d^2 r_1 \int d^2 r_2 \left[
\frac{1}{2} e \varphi (\tau, \mathbf{r}_1)
V_c^{-1} (\mathbf{r}_1, \mathbf{r}_2)  e \varphi (\tau, \mathbf{r}_2)
+ i e \varphi(\tau, \mathbf{r}_1) n \delta(\mathbf{r}_1 - \mathbf{r}_2)
\right] +
T \mbox{Tr} \sum_{n=1}^{\infty}
\frac{1}{n} ({\cal G} \Sigma)^{n}\,.
%\right|_{\partial \Delta/\partial \mathbf{R}=0}\,.
\end{split}
\end{equation}
This way of deriving the effective action has many advantages. The main among them
is that the gauge invariant combinations
$\hbar \partial_{\tau} \theta/2 - e \varphi - e A_{0}$ and
$\hbar \nabla \theta/2  - e/c \mathbf{A}$ are explicitly present
during all stages of the derivation \cite{foot3}.
This property is obviously related to the introduction of the phase via the
gauge transformation Eq.~(\ref{gauge}). There is no need because of this
to keep the external electromagnetic field $(A_0, \mathbf{A})$
during the intermediate stages of the derivation since it can be easily
restored following the above mentioned prescription which in the
frequency-momentum space are
\begin{equation}
\label{restore.field}
i \Omega_n \theta(K)  \to i \Omega_n \theta(K) + 2 e A_{0}(K);
\qquad
i K_{\alpha} \theta(K)  \to  i K_{\alpha} \theta(K) -
\frac{2 e}{c} A_{\alpha}(K).
\end{equation}
Differentiating with respect to this source field we will derive
physical correlation functions (see the discussion in 
\cite{Randeria.action}) in what follows. 
It has to be stressed that the {\it minimal coupling} prescription
(\ref{restore.field}) does not guaranty itself the gauge independence
of the final result. The gauge independent treatment of the transformations
(\ref{gauge}) and (\ref{Phi.modulus.approx}) for $d$-wave pairing
is, in particular, one of the complications \cite{foot2}.
%Another known complication is that the minimal coupling (\ref{restore.field})
%cannot be used unless the effective Lagrangian is local. Thus even for 
%$s$-wave pairing there is a problem how to include electromagnetism when
%the Landau damping terms are present \cite{Aitchison.2000}.

In the previous studies of the CG mode
\cite{Kulik.1981,Takada.1988,Takada.1997,Takada.2000} the phase field
was introduced using the expansion of the ordering field $\Phi(x)$
around the equilibrium value $\Delta$ via
$\Phi(x) = \Delta + \Phi_{1}(x) + i \Phi_{2}(x)$
and associating the fields $\Phi_{1}(x)$ with the amplitude and
$\Phi_2(x)$ (or to be more precise $\Phi_2(x) / \Delta_0$) with
the phase fluctuations. Although, as will be discussed below,
the final result obtained in the both methods agrees,
the present method of the investigation of the CG mode is more transparent
because it explicitly uses the gauge independent combinations
of the fields over the whole derivation. For example, one can easily recognize
that $\Sigma$ contains a gauge independent Cooper pair chemical potential
$\dot{ \theta}/2 - e \varphi$ \cite{book.1986} which in other approaches
has to be collected from the different parts of the equations.
Other advantages of the present approach will be discussed in the subsequent
sections where the effective action is presented.

To consider the phase and charge fluctuations at the Gaussian level
it is sufficient to include only the terms with $n=1,2$ in the infinite series in
Eq.~(\ref{Omega.Kinetic.phase}). Finally, we rewrite the pair susceptibility
Eq.~(\ref{pair.susceptibility}) in the new variables
\begin{equation}
\label{pair.susceptibility.full}
\eta(\tau, \mathbf{R}_1 - \mathbf{R}_2) = -
\int \Delta \mathcal{D} \Delta \mathcal{D} \theta
\mathcal{D} \varphi
\Delta(\tau, \mathbf{R}_1, \mathbf{0}) \exp[i\theta (\tau, \mathbf{R}_1)]
\Delta (0, \mathbf{R}_2, \mathbf{0})  \exp[-i\theta (\tau, \mathbf{R}_2)]
\exp[ - \beta \Omega\{ \Delta, \partial \theta, \varphi \}] \,.
\end{equation}
As was mentioned above, we are interested only in the phase
fluctuations structure factor neglecting the presence of the
amplitude fluctuations. This implies that one can use a saddle
point approximation for $\Delta$, so that omitting unimportant
constant in the expansion of the exponent in
Eq.~(\ref{pair.susceptibility.full}) one arrives at
\begin{equation}
\label{pair.susceptibility.phase}
\eta(\tau, \mathbf{R}_1 - \mathbf{R}_2) = - \frac{\Delta_0^2}{4}
\int  \mathcal{D} \theta
\mathcal{D} \varphi
\theta (\tau, \mathbf{R}_1) \theta (\tau, \mathbf{R}_2)
\exp[ - \beta \Omega_{kin}\{\partial \theta, \varphi \}] \,.
\end{equation}
Writing Eq.~(\ref{pair.susceptibility.phase}) we also expanded the exponents which
were present in Eq.~(\ref{pair.susceptibility.full}),
because there are no free vortices in the system for $T < T_c$ and the
multivalued character of the phase is irrelevant. This approximated form
of the pair susceptibility is equivalent to the expressions for the susceptibility
used in Refs.~\cite{Kulik.1981,Takada.1988,Takada.1997,Takada.2000}.
Expanding the exponents we neglect a widening of the structure factor peaks
which is related to the absence of the long-range order in 2D (CMWH theorem).
It is known, for example, from the analysis of the dynamic structure factor of lattices
\cite{Mikeska,Weling} that $\delta$-function phonon resonance obtained in
3D harmonic crystals in 2D case is converted to the power law singularity
\begin{equation}
\label{peak}
S(\omega, \mathbf{q}) \sim \frac{f(\alpha)}{|\omega^2 - v^2 q^2|^{1- \alpha(T)}},
\end{equation}
where $f(\alpha)$ is a function of $\alpha$ which goes to zero as  $\alpha \to 0$,
so that in this limit the structure factor transforms to $\delta$-function.
Since for low temperatures $\alpha(T) \ll 1$ we
may safely neglect this effect of widening
because it does not move the position of the peak and we are primarily
interested in the temperatures $T \ll T_{c}$.

\section{General form of the effective action}
\label{sec:Omega}

In this section we present the effective potential
$\Omega_{kin}\{ \theta, \varphi \}$ and
discuss the term which leads to the appearance of the CG mode. We also
derive the effective ``phase only'' and ``electric potential only'' actions
integrating out the electric potential $\varphi$ and the phase $\theta$, respectively.

\subsection{The effective action and polarization functions}

Calculating the terms with $n =1,2$ in Eq.~(\ref{Omega.Kinetic.phase})
(see e.g. \cite{Aitchison.2000,we.LD}) one arrives at
\begin{equation}
\label{phase.electric.action}
\begin{split}
\beta \Omega_{kin} \{\theta, \varphi \} =
\frac{T}{8} \sum_{n = - \infty}^{\infty}
& \int \frac{d {\bf K}}{(2 \pi)^2}  \left[
 \varphi(-K) 4 e^2 V_{c}^{-1}({\bf K}) \varphi(K) +
(i\Omega_n \theta(-K) - 2 e  \varphi(-K)) \Pi_{33}(K)
(i \Omega_n \theta(K) + 2 e \varphi(K)) \right. \\
& + \theta(-K) [\Lambda_{0}^{\alpha \beta} + \Pi_{00}^{\alpha \beta}(K)]
K_{\alpha} K_{\beta} \theta(K) \\
&  \left. + \theta(-K) (-K_{\alpha}) \Pi_{30}^{\alpha}(K)
(i \Omega_n \theta(K)+ 2 e\varphi(K)) +
(i \Omega_n \theta(-K)- 2 e \varphi(-K)) (-K_{\alpha}) \Pi_{03}^{\alpha}(K)
\theta(K) \right],
\end{split}
\end{equation}
where we introduced short-hand notations $K = (i \Omega_n, {\bf K})$
with $\mathbf{K}$ being 2D vector (summation over dummy indices $\alpha, \beta =1,2$
is implied).
In  Eq.~(\ref{phase.electric.action})
the current-current polarization function, $\Pi_{00}$  is
\begin{equation}\label{Pi00}
\Pi_{00}^{\alpha \beta}(i\Omega_{n}, {\bf K}) \equiv
T \sum_{l = - \infty}^{\infty} \int \frac{d^2 k}{(2 \pi)^2}\,
\pi_{00}(i\Omega_{n},{\bf K}; i \omega_{l}, {\bf k})
v_{F \alpha}({\bf k}) v_{F \beta}({\bf k})\,
\end{equation}
with the Fermi velocity
$v_{F \alpha}(\mathbf{k})  =
\partial \xi(\mathbf{k})/\partial k_{\alpha}$;
the density-density  polarization function, $\Pi_{33}$ is
\begin{equation}\label{Pi33}
\Pi_{33}(i\Omega_{n}, {\bf K}) \equiv
T \sum_{l = - \infty}^{\infty} \int \frac{d^2 k}{(2 \pi)^2}\,
\pi_{33}(i\Omega_{n},{\bf K}; i \omega_{l}, {\bf k})\,
\end{equation}
and the density-current polarization function, $\Pi_{03}^{\alpha}$ is
\begin{equation}
\label{Pi03}
\Pi_{03}^{\alpha}(i\Omega_{n}, {\bf K}) \equiv
T \sum_{l = - \infty}^{\infty} \int \frac{d^2 k}{(2 \pi)^2}\,
\pi_{03}(i\Omega_{n},{\bf K}; i \omega_{l}, {\bf k})
v_{F \alpha}({\bf k}) \,.
\end{equation}
$\pi_{ij}$ in Eqs.~(\ref{Pi00}) - (\ref{Pi03}) is given by
\begin{equation}\label{pi}
\pi_{ij}(i\Omega_{n},{\bf K}; i \omega_{l}, {\bf k}) \equiv
\mbox{tr} [\mathcal{G} (i \omega_{l} + i\Omega_{n}, {\bf k} + {\bf K}/2)
\tau_{i} \mathcal{G} (i \omega_{l}, {\bf k} - {\bf K}/2) \tau_{j}]\,,
\qquad (\tau_{0} \equiv \hat{I})\,.
\end{equation}
and $\Lambda_{0}^{\alpha \beta}$ in Eq.~(\ref{phase.electric.action})
is the first order contribution in the superfluid stiffness:
\begin{equation}\label{Lambda0}
\Lambda_{0}^{\alpha \beta} = \int
\frac{d^2 k}{(2 \pi)^2} n(\mathbf{k})  m_{\alpha \beta}^{-1} (\mathbf{k}),
\qquad m_{\alpha \beta}^{-1}(\mathbf{k})\equiv
\partial^2 \xi(\mathbf{k})/\partial k_{\alpha} \partial k_{\beta}
\end{equation}
with
\begin{equation}\label{n(k)}
n(\mathbf{k}) = 1 - \frac{\xi(\mathbf{k})}{E(\mathbf{k})} \tanh
\frac{E(\mathbf{k})}{2T}\,, \qquad E(\mathbf{k}) =
\sqrt{\xi^2(\mathbf{k})+\Delta^2(\mathbf{k})}\,.
\end{equation}
Writing Eq.~(\ref{phase.electric.action}) we omitted the linear time derivative term
(see e.g. \cite{Randeria.action,we.LD}) which is irrelevant for the present
analysis.

The general expressions for the polarizations (\ref{Pi00}) - (\ref{Pi03})
are given in Appendix~\ref{sec:A} (see also \cite{we.LD}) and calculated
analytically for 2D clean $d$-wave superconductor in Appendix~\ref{sec:C}
(a brief discussion of the nodal approximation used to calculate
these polarizations and the transformation to the global coordinate system
are given in Appendices \ref{sec:B} and \ref{sec:D}, respectively).

\begin{figure}[h]
\centering{
\includegraphics[width=8cm]{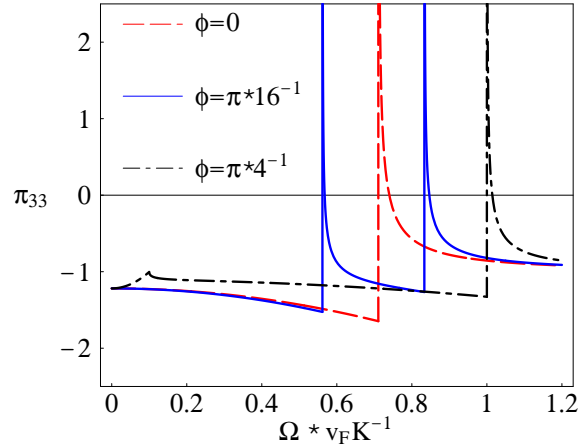}}
\caption{Real part of $\Pi_{33}(\Omega, \mathbf{K})$ (or more precisely
${\tilde \Pi}_{33}(\Omega, \mathbf{K}) = (\pi/m) \Pi_{33}(\Omega, \mathbf{K})
$, see Eq.~(\ref{pi_{33}})).
We put $T = 0.5 T_c$ ($T_c$ and the angle $\phi$ are defined
in Appendices~\ref{sec:C} and~\ref{sec:D}, respectively)
and $\alpha_{D} = 10$.}
\label{fig:1}
\end{figure}

As an example we show in Fig.~\ref{fig:1} the real part of the
density-density polarization function, $\Pi_{33}(\Omega,
\mathbf{K})$ (this function is in fact just the Lindhard's
function for the superconducting state) which is given by
Eq.~(\ref{Pi_{33}.final}) (and its dimensionless form by
Eqs.~(\ref{pi_{33}.definition}) and (\ref{pi_{33}})) as a function
of  $\Omega/v_{F} K$ for the different  directions of
$\mathbf{K}$. The angle $\phi$ determines the direction of
$\mathbf{K}$ with respect to the $(10)$-direction in a such way
that $\phi = \pi/4$ corresponds to the nodal direction (see also
Eq.~(\ref{transform}) and the explanation in
Appendix~\ref{sec:D}). Comparing this figure with Fig.~7 from
\cite{Takada.2000} (our  definition of $\phi$ is equivalent to the
definition of the angle $\theta_{\mathbf{q}}$ used in
\cite{Takada.2000}), which was obtained by numerical integration
of Eq.~(\ref{Pi.general}), one can see that our analytical
expression (\ref{Pi_{33}.final}) gives essentially the same
result. In particular, Fig.~\ref{fig:1} shows that $\Pi_{33}$ has
a peak at $\Omega/v_{F} K = \cos (\phi - \pi/4)$. Furthermore, as
shown, for instance, for $\phi = \pi/16$ there is a lower peak in
$\mbox{Re} \Pi_{33}(\Omega, \mathbf{K})$ at $\Omega/v_{F} K =
\cos(\phi - \pi/4 + \pi/2)$. Note that the cases $\phi =0$ and
$\phi = \pi/4$ are ``degenerate'' because for $\phi =0$ the lower
peak coincides with the upper one and for $\phi = \pi/4$ the lower
peak is at $\Omega/v_{F} K =0$ (see also Fig.~\ref{fig:5}, where
the  case $\phi = 0.2, 0.23 \pi$ is shown).

In \cite{Takada.2000} the origin of these peaks was related to the
gap nodes which can be regarded as two ``one-dimensional normal
state electronic bands'' toward $\phi = \pm \pi/4$. These ``normal
bands'' are able to screen out the Coulomb interaction in  certain
regions of the Fermi surface even for $T \ll T_c$ and this
screening along with the presence of $\Pi_{03}^{\alpha}$ will make
the appearance of the CG mode possible. Substituting
Eq.~(\ref{transform})  in the analytical expression
(\ref{Pi_{33}.final})  it is indeed easy to see that these peaks
are due to the square root singularities of $\Pi_{33}$. These are
the same singularities which are present in 2D normal state
Lindhard's function due to the lowered dimensionality of the
momentum integration \cite{Tsvelik}, but since $d$-wave
superconducting state is considered the position of these
singularities does depend on the direction of $\mathbf{K}$ with
respect to the Fermi surface. Finally we note that these
singularities in $\mbox{Re} \Pi_{33}$ at $\Omega = v_{F} K
\cos(\phi - \pi/4) + 0$ is accompanied by the singularity in
$\mbox{Im} \Pi_{33}$ at $\Omega = v_{F} K \cos(\phi - \pi/4) - 0$
which was considered in \cite{we.LD}.

The effective potential Eq.~(\ref{phase.electric.action})
becomes more tractable in the matrix form
\begin{equation}\label{matrix.potential}
\beta \Omega_{kin}\{\theta, \varphi \} = \frac{T}{8}
\sum_{n = -\infty}^{\infty} \int \frac{d {\bf K}}{(2 \pi)^2}
\begin{bmatrix}
\theta(-K) \, \,   & e \varphi(-K)
\end{bmatrix}
\mathcal{M}^{-1}
\begin{bmatrix}
\theta(K) \\ e \varphi(K)
\end{bmatrix}\,,
\end{equation}
where
\begin{equation}\label{M}
\mathcal{M}^{-1} =
\begin{bmatrix}
-\Omega_n^2 \Pi_{33}(K) + \Lambda^{\alpha \beta}(K) K_{\alpha} K_{\beta}-
i \Omega_n K_{\alpha} \Pi_{03}^{\alpha}(K) - i \Omega_n K_{\alpha} \Pi_{30}^{\alpha}(K) &
2 i \Omega_n \Pi_{33}(K) - 2 K_{\alpha} \Pi_{30}^{\alpha}(K) \\
-2 i \Omega_n \Pi_{33}(K) + 2 K_{\alpha} \Pi_{30}^{\alpha}(K) &
4 (-\Pi_{33}(K) + V_{c}^{-1}({\bf K}))
\end{bmatrix}
\end{equation}
with the bare (unrenormalized by the phase fluctuations) superfluid
stiffness $\Lambda^{\alpha \beta} = \Lambda_{0}^{\alpha \beta} +
\Pi_{00}^{\alpha \beta}(K)$.

\subsection{Comparison with other approaches}
\label{sec:comparison}

Let us  compare our effective action
Eqs.~(\ref{matrix.potential}), (\ref{M}) with the action obtained
in \cite{Randeria.action} and see the differences between the
present and previous
\cite{Kulik.1981,Takada.1988,Takada.1997,Takada.2000} approaches.
As one can notice, the only difference between
Eqs.~(\ref{matrix.potential}), (\ref{M}) and Eqs.~(25), (26) in
\cite{Randeria.action} is due to the presence of the
density-current polarization function, $\Pi_{03}^{\alpha}$. It is
a general belief that this term has to be zero due to the
``symmetry arguments'' \cite{Nagaosa.book}. However, as shown in
\cite{Aitchison.2000} (see also \cite{we.LD}, where the $d$-wave
case is considered) this correlator has to be taken into account
to obtain the correct expressions for the {\em Landau terms\/} of
the effective action. This is the term which nontrivially couples phase
and density fluctuations and makes the CG mode possible in the
present approach.
%Furthermore, the presence of this term is required
%to maintain the gauge invariance for $\Omega \neq 0$.

From this point of view, the role of $\Pi_{03}^{\alpha}$ is the same as
the role of the phase-charge coupling
\begin{equation}\label{Pi23}
\Pi_{23}(i\Omega_{n}, {\bf K}) \equiv
T \sum_{l = - \infty}^{\infty} \int \frac{d^2 k}{(2 \pi)^2}\,
\pi_{23}(i\Omega_{n},{\bf K}; i \omega_{l}, {\bf k}) \gamma(\mathbf{k})\,,
\qquad \gamma(\mathbf{k}) \equiv \frac{2 \Delta(\mathbf{k})}{\Delta_0},
\end{equation}
with $\pi_{23}$ given by Eq.~(\ref{pi}) in the approach of Kulik
et al. \cite{Kulik.1981} and Takada with coauthors
\cite{Takada.1988,Takada.1997,Takada.2000}. Note that $\Pi_{23} = - \Pi_{32}$,
while $\Pi_{03}^{\alpha} = \Pi_{30}^{\alpha}$.

It is interesting that  techniques
essentially similar with \cite{Kulik.1981,Takada.1988,Takada.1997,Takada.2000}
have been used in \cite{Varlamov.1986} and \cite{Smith.1995,Smith.2000}
to consider suppression of the critical temperature in disordered
superconductors. In \cite{Smith.1995,Smith.2000} both amplitude and
phase fluctuations were taken into account and to consider the influence
of nonmagnetic impurities the electronic Green's functions had
$4 \times 4$ matrix structure. The main difference between
\cite{Kulik.1981,Takada.1988,Takada.1997,Takada.2000,Varlamov.1986,Smith.1995,Smith.2000},
nevertheless, remained the same, viz.
the order parameter phase was expressed via the operator
$O_2 = \Psi^{\dagger} \tau_2 \Psi$, as summarized
in Table~II of \cite{Smith.1995}. Thus $\tau_2$ also enters the
phase-density correlator, $\Pi_{23}$ in the notations of
\cite{Kulik.1981,Takada.1988,Takada.1997,Takada.2000}
(or $\Pi_{\phi \rho}$ in the notations of \cite{Smith.1995,Smith.2000}).

In our opinion, the physical meaning
of $\Pi_{23}$ is, however, more obscure than that of $\Pi_{03}^{\alpha}$.
Indeed, since $\Pi_{23}$ is expressed via the Pauli matrix $\tau_2$, so that
it seems like the phase itself is a dynamical variable on its own, while
physically meaningful are only the space and time derivatives of the
phase.

These derivatives can only enter into the formal expressions as a
current via $\hat I$ and as a density via $\tau_{3}$ matrices,
respectively. This property is obviously present in the definition
Eq.~(\ref{Pi03}) of $\Pi_{03}^{\alpha}$ which thus has the more
clear meaning of a density-current polarization function.

%??? I should further develop this consideration here or below.

Another important difference between the present and previous
\cite{Kulik.1981,Takada.1988,Takada.1997,Takada.2000,Varlamov.1986,Smith.1995,Smith.2000}
approaches is that the present derivation does not need an
explicit use of the gap equation for $\Delta$. For example, in
\cite{Kulik.1981} the charge conservation follows from the
explicit use of the gap equation, while in the present approach as
we will discuss later, the charge conservation is already built in
the phase dynamics itself.

The fact that the present approach does not rely on the particular
form of the gap equation is more convenient for modeling HTSC,
where the gap  does not close at the critical temperature $T_c$,
so that the equation $\Delta(T_c^0) = 0$ gives only the mean-field
transition temperature, $T_c^0$. Thus another definition of the
true critical temperature, $T_c$ is necessary. As recently
discussed in \cite{Benfatto.dissipation} (see also
\cite{Loktev.review}), it is reasonable for HTSC to estimate $T_c$
as the temperature of the Berezinskii-Kosterlitz-Thouless
transition (see Eq.~(\ref{Tc}) in Appendix~\ref{sec:C}).

Finally, it is worth to mention here the recent papers \cite{Gusynin}, where
the approach very similar to that of the  present paper was employed to study the
CG mode in the model of color superconducting quark matter. One of the advantages
of \cite{Gusynin} is that it treats the problem in an explicitly gauge invariant way,
while here, the Coulomb gauge is already imposed in writing the Hamiltonian
Eq.~(\ref{Hamilton}) and then, when necessary, the gauge independence of
the results obtained is discussed in a more intuitive way.
In general, however, to prove the gauge independence of the results
an arbitrary gauge has to be considered, to show that the physical
observables do not depend on the gauge fixing parameter as
done in \cite{Gusynin} (see also \cite{Aitchison.1984}, where the gauge
invariance of the physical quantities calculated using the 
$T =0$ effective potential is discussed). Following this route one can obtain
the ``relativistic'' (see the second paper in \cite{Gusynin}
for the details of the proof) generalization of 
Eq.~(\ref{phase.electric.action}) which 
contains a gauge fixing parameter $\lambda$ 
\begin{equation}\label{V.P.}
\begin{split}
\beta \Omega_{kin} \{\theta, A_{\mu} \} =  
& T \sum_{K} \frac{1}{8} 
\left\{ \frac{1}{\pi}\left[ A_{\mu}(-K)
(K^2 \delta_{\mu \nu} - K_{\mu} K_{\nu}) A_{\nu} (K)
+ \frac{1}{\lambda} A_{\mu}(-K) K_{\mu} K_{\nu}  A_{\nu} (K) \right] \right. \\
& \left. + (-i K_{\mu} \theta(-K) - 2 e A_{\mu}(-K)) \Pi_{\mu \nu}(K)
(i K_{\nu} \theta(K) - 2 e A_{\nu}(K)) \right\}, 
\end{split}
\end{equation}
with $K_{\mu} \equiv (\Omega_n, \mathbf{K})$,
$K^2 = K_{\mu} K_{\mu} = \Omega_n^{2} + \mathbf{K}^2$,
$\mu,\nu= 0,1,2$ and $A_0(K) = - \varphi (K)$ and $c=1$.
Note that in Eq.~(\ref{V.P.}) we have the whole electromagnetic potential $A_{\mu}$
instead of the Coulomb component present in Eq.~(\ref{phase.electric.action}).
The polarization tensor $\Pi_{\mu \nu} (K)$ is obviously related to the
polarizations used in Eq.~(\ref{phase.electric.action}).
The question of gauge independence (or dependence) can be addressed considering
how the calculated quantities depend on $\lambda$.

\subsection{Effective actions for the phase and electric potential}

Integrating out $\varphi$ and $\theta$ from Eq.~(\ref{matrix.potential})
one can obtain, respectively
\begin{equation}\label{theta}
\beta \Omega_{kin}\{\theta\} =
\frac{T}{8} \sum_{n= - \infty}^{\infty} \int \frac{d {\bf K}}{(2 \pi)^2}
\theta(-K) \mathcal{M}_{\theta}^{-1} (i \Omega_n, \mathbf{K})
\theta(K),
\qquad
\mathcal{M}_{\theta}^{-1} = M_{11}^{-1} - M_{22} M_{12}^{-1} M_{21}^{-1}
\end{equation}
and
\begin{equation}\label{varphi}
\beta \Omega_{kin}\{\varphi\} =
\frac{T}{8}  \sum_{n= - \infty}^{\infty} \int \frac{d {\bf K}}{(2 \pi)^2}
\varphi(-K) \mathcal{M}_{\varphi}^{-1} (i \Omega_n, \mathbf{K})
\varphi(K),
\qquad \mathcal{M}_{\varphi}^{-1} =
e^{2}(M_{22}^{-1} - M_{11} M_{12}^{-1} M_{21}^{-1}) \,.
\end{equation}
It is evident that $\mathcal{M}_{\varphi}^{-1} =
e^2 M_{22}^{-1} M_{11} \mathcal{M}_{\theta}^{-1}
=e^2 M_{22}^{-1} \mbox{det} \mathcal{M}^{-1}$, so that if
$M_{11}^{-1}, M_{22}^{-1} \ne 0$ the equations
$\mathcal{M}_{\theta}^{-1} = \mathcal{M}_{\varphi}^{-1} =
\mbox{det} \mathcal{M}^{-1} = 0$ are equivalent.

Using Eqs.~(\ref{pair.susceptibility.phase}) and (\ref{varphi})
it is straightforward to obtain that the structure factor
Eq.~(\ref{structure.factor}) is given by
\begin{equation}
\label{structure.factor.final}
S(\Omega, \mathbf{K}) = \frac{T}{\Omega} \frac{\Delta_0^2}{8} \mbox{Im}
\mathcal{M}_{\theta} (\Omega + i 0, \mathbf{K})\,.
\end{equation}

\section{Phase excitations in the charged system in the  absence
of the density-current coupling}
\label{sec:standard}

\subsection{Equation for the collective phase excitations}

Let us assume for a moment that there is no density-current
coupling ($\Pi_{03}^{\alpha}=\Pi_{30}^{\alpha}= 0$)
and discuss briefly the collective excitations which follow
from Eqs.~(\ref{M}), (\ref{theta}) and (\ref{varphi}).
As mentioned above the matrix $\mathcal{M}^{-1}$ reduces in this
case to the known expression \cite{Randeria.action}. Therefore,
it is not surprising that the ``phase only'' action
Eq.~(\ref{theta}) takes the form
\begin{equation}\label{phase.plasmon}
\mathcal{M}_{\theta}^{-1}(i\Omega_n, {\bf K}) = -\Omega_n^{2}
\mathbf{\bar{\Pi}}_{33}  +
\Lambda^{\alpha \beta} K_{\alpha} K_{\beta},
\qquad
\mathbf{\bar{\Pi}}_{33} = \frac{\Pi_{33}(K)}{1 - \Pi_{33}(K) V_{c}({\bf K})}\,,
\end{equation}
which coincides with the corresponding expression in
\cite{Randeria.action} (see also \cite{Nagaosa.book}).
The dispersion law of the collective phase modes is defined by the
equation
\begin{equation}\label{phase.modes}
\mathcal{M}_{\theta}^{-1}(i\Omega_n \to \Omega +i0,{\bf K}) =0.
\end{equation}
This equation can also be regarded as a direct consequence of the charge
conservation
\begin{equation}
\label{conservation}
\frac{\partial \rho(t, \mathbf{r})}{\partial t} +
\nabla \cdot \mathbf{j}(t, \mathbf{r}) =0\,,
\end{equation}
where the current and charge density are defined via
\begin{equation}
\label{current-density.general}
\mathbf{j} = -c \frac{\delta \Omega_{kin}\{\theta,A \}}{\delta \mathbf{A}},
\qquad \rho = \frac{\delta \Omega_{kin}\{\theta,A \}}{\delta A_0}\,,
\end{equation}
where the electromagnetic field $A$ in $\Omega_{kin}$ was restored
using the rule Eq.~(\ref{restore.field}). Evaluating
Eq.~(\ref{current-density.general}) one arrives at
Eq.~(\ref{current-density}) with $\Pi_{03}^{\alpha} =0$, so that
Eq.~(\ref{phase.modes}) is indeed recovered. There is, in fact, no
surprise in the link between Eqs.~(\ref{conservation}) and
(\ref{phase.modes}) which is just the consequence of the way how
we introduced the phase $\theta$ in Eq.~(\ref{gauge}).

As discussed in \cite{Randeria.action,Nagaosa.book} the only solution
of this equation for $\mathbf{K} \to 0$ is the plasma mode.
Using as an example the 3D form of the Coulomb potential and
assuming that $\Lambda^{\alpha \beta} = \Lambda \delta_{\alpha \beta}$
which is valid for isotropic system with
$m_{\alpha \beta} = m \delta_{\alpha \beta}$, one obtains
from this equation that the plasma frequency,
$\omega_p = \sqrt{4 \pi \Lambda e^2}$ for the limit $\mathbf{K} \to 0$.
The expression for $\omega_p$ can be reduced to the standard
$\sqrt{4 \pi n e^2/m}$ if one uses the superfluid stiffness
$\Lambda = n/m$ obtained for the continuum model with $s$-wave pairing.

It is clear that for the plasma mode $\Omega_{p}(\mathbf{K})/K \to \infty$
as $\mathbf{K} \to 0$. This property remains valid even if 2D Coulomb
potential is used and it makes plasmons different from any sound mode with
$\Omega /K \to \mbox{const}$ as $\mathbf{K} \to 0$.
For example, if a neutral superconductor were considered the polarization
$\mathbf{\bar{\Pi}}_{33}$ in Eq.~(\ref{phase.plasmon}) would
be replaced by $\Pi_{33}$ and  the solution of
Eq.~(\ref{phase.modes}) for $\mathbf{K} \to 0$
is the sound-like BA mode and its velocity
is given by Eq.~(\ref{v}).

%For $T \neq 0$ this mode becomes decaying, so that in general the equation
%$ \mathcal{M}_{\theta}^{-1}(i\Omega_n \to \Omega +i0,{\bf K}) =0$
%has to be considered.

\subsection{Phase excitations via electric potential propagator and
gauge independent density-density and current-current correlators}

It is instructive also to look at the form $\mathcal{M}_{\varphi}$
for the electric potential $\varphi$ which is
\begin{equation}\label{varphi.standard}
\begin{split}
\mathcal{M}_{\varphi}^{-1}(i\Omega_n, {\bf K})
%&  = \frac{4 e^2
%[-\Pi_{33} \Lambda^{\alpha \beta} K_{\alpha} K_{\beta} -
%V_{c}^{-1}({\bf K}) \Omega_n^{2} \Pi_{33} +
%V_{c}^{-1}({\bf K}) \Lambda^{\alpha \beta} K_{\alpha} K_{\beta}]}
%{-\Omega_n^2 \Pi_{33} + \Lambda^{\alpha \beta} K_{\alpha} K_{\beta}} \\
%&
= 4 e^2 \left(
\frac{-\Pi_{33} \Lambda^{\alpha \beta} K_{\alpha} K_{\beta}}
{-\Omega_n^2 \Pi_{33} + \Lambda^{\alpha \beta} K_{\alpha} K_{\beta}}
+ V_c^{-1}(\mathbf{K}) \right)\,.
\end{split}
\end{equation}
Considering the same example of isotropic system with 3D Coulomb
potential Eq.~(\ref{varphi.standard}) in the limit $\mathbf{K} \to 0$
and  $\Omega/K \to \infty$
can be reduced to the known expression (see e.g. \cite{Nagaosa.book})
\begin{equation}
\mathcal{M}_{\varphi}^{-1}(i\Omega_n, {\bf K}) =
\frac{\mathbf{K}^2}{\pi} \left(1 + \frac{\omega_p^2}{\Omega_n^2} \right)\,.
\end{equation}
It is obvious that the discussed above plasma mode can be also seen in
$\mathcal{M}_{\varphi}$. Indeed after the analytical continuation
$i \Omega_n \to \Omega + i0$, $\mathcal{M}_{\varphi}(\Omega, \mathbf{K})$
acquires a pole at $\Omega = \omega_p$.

It is also useful to evaluate gauge independent density-density and
current-current correlators which are defined as
\begin{equation}
\label{cor.def}
\chi(i \Omega_n, {\bf K}) =
- \frac{\delta^2 T \ln Z\{ A\}}{\delta A_{0} (-K)\delta A_{0}(K)},
\qquad
\chi^{\alpha \beta}(i \Omega_n, {\bf K}) = -
\frac{\delta^2 T \ln Z\{ A\}}
{\delta A_{\alpha} (-K)\delta A_{\beta}(K)},
\end{equation}
where
\begin{equation}
Z\{A \} = \int \mathcal{D} \theta
\mathcal{D} \varphi \exp[- \beta \Omega_{kin}\{\theta, \varphi, A\}]
\end{equation}
with the external field $(A_0, \mathbf{A})$ restored using
the rule Eq.~(\ref{restore.field}).
Then we arrive at the standard expressions \cite{Randeria.action}
\begin{equation}\label{density.cor.standard}
\chi(i \Omega_n, {\bf K}) = -
\frac{e^2 \mathbf{\bar{\Pi}}_{33} \Lambda^{\alpha \beta}K_{\alpha} K_{\beta}}{
\Lambda^{\alpha \beta}K_{\alpha} K_{\beta} - \mathbf{\bar{\Pi}}_{33} \Omega_n^{2}}
\end{equation}
and
\begin{equation}\label{current.cor.standard}
\chi^{\alpha \beta}(i \Omega_n, {\bf K}) = e^2 \left(
\Lambda^{\alpha \beta}
- \frac{\Lambda^{\alpha \mu} \Lambda^{\nu \beta} K_{\mu} K_{\nu}}
{- \Omega_n^2 \mathbf{\bar{\Pi}}_{33} +
\Lambda^{\alpha \beta} K_{\alpha} K_{\beta}} \right) \,.
\end{equation}

Again assuming that $\Lambda^{\alpha \beta} = \Lambda \delta_{\alpha \beta}$
one can reduce Eq.~(\ref{current.cor.standard}) to the known form of the
current-current correlator
\begin{equation}\label{current.cor.}
\chi^{\alpha \beta}(i \Omega_n, {\bf K}) = e^2 \Lambda
\frac{-\Omega_n^2 \mathbf{\bar{\Pi}}_{33} \delta_{\alpha \beta}
+ \Lambda K^2 \left(\delta_{\alpha \beta} -
\frac{K_{\alpha} K_{\beta}}{K^2} \right)}
{- \Omega_n^2 \mathbf{\bar{\Pi}}_{33} + \Lambda K^2}.
\end{equation}
The difference between gauge independent
Eq.~(\ref{density.cor.standard}) and
Eq.~(\ref{current.cor.standard}) and gauge dependent
$\mathbf{\bar{\Pi}}_{33}$,  $\Lambda^{\alpha \beta}$ in
Eq.~(\ref{phase.plasmon}) was recently discussed in
\cite{Randeria.action}. The link considered above  between charge
conservation Eq.~(\ref{conservation}) and Eq.~(\ref{phase.modes})
shows that the solutions of Eq.~(\ref{phase.modes}) are gauge
independent, even though $\mathbf{\bar{\Pi}}_{33}$,
$\Lambda^{\alpha \beta}$ are gauge dependent. 
At the more formal level it can be argued that
Eqs.~(\ref{density.cor.standard}) and (\ref{current.cor.standard})
are gauge independent because even if one derives them starting from
Eq.~(\ref{V.P.}) the gauge fixing parameter $\lambda$ {\it does not enter}
the final result and one obtains the same expressions.

The putting this in another way, one can say that the position of zeros of
$\mathcal{M}_{\theta}^{-1}(\Omega, \mathbf{K})$ is gauge
independent, because these zeros coincide with the poles of the
gauge independent $\chi(\Omega, \mathbf{K})$. As we will see in
Sec.~\ref{sec:correlators}, density-current coupling modifies both
Eqs.~(\ref{density.cor.standard}) and
(\ref{current.cor.standard}), nevertheless the general argument
about zeros of $\mathcal{M}_{\theta}^{-1}(\Omega, \mathbf{K})$
(poles of $\mathcal{M}_{\theta}(\Omega, \mathbf{K})$) remains
valid.

\section{Phase excitations in the presence of the density-current
coupling}
\label{sec:main}

Here we generalize all expressions from the previous section
for nonzero $\Pi_{30}^{\alpha}$.
In particular, Eq.~(\ref{phase.plasmon}) becomes
\begin{equation}\label{theta.cor.new}
\begin{split}
& \mathcal{M}_{\theta}^{-1}(i\Omega_n, {\bf K})
%&  =
%\frac{\Lambda^{\alpha \beta} K_{\alpha} K_{\beta}
%[- \Pi_{33} + V_{c}^{-1}({\bf K})]  -
%\Omega_n^{2} \Pi_{33} V_{c}^{-1}({\bf K}) -
%V_{c}^{-1}({\bf K}) i\Omega_n K_{\alpha} (\Pi_{03}^{\alpha}+ \Pi_{30}^{\alpha})
%+ K_{\alpha}K_{\beta} \Pi_{30}^{\alpha} \Pi_{03}^{\beta}}
%{-\Pi_{33} + V_c^{-1}({\bf K})}  \\
  = - \Omega_n^2 \mathbf{\bar{\Pi}}_{33}  -
i \Omega_n K_{\alpha}( \mathbf{\bar{\Pi}}_{03}^{\alpha}(K)+
\mathbf{\bar{\Pi}}_{30}^{\alpha}(K))
+ \mathbf{\bar{ \Lambda}}^{\alpha \beta} K_{\alpha} K_{\beta}, \\
& \mathbf{\bar{\Pi}}_{03}^{\alpha}(K) \equiv
\frac{\Pi_{03}^{\alpha}(K)}{1 - \Pi_{33}(K) V_{c}({\bf K})}, \qquad
\mathbf{\bar{\Lambda}}^{\alpha \beta} \equiv
\Lambda^{\alpha \beta} +
\frac{\Pi_{03}^{\alpha}(K) \Pi_{30}^{\beta}(K)}{- \Pi_{33}(K) + V_c^{-1}({\bf K})}.
\end{split}
\end{equation}
The dispersion law for all collective phase excitations
is still defined by Eq.~(\ref{phase.modes}), but with
$\mathcal{M}_{\theta}$ given by Eq.~(\ref{theta.cor.new}).
It has to be pointed out that the interpretation of Eqs.~(\ref{phase.modes})
and  (\ref{theta.cor.new}) as the charge conservation law Eq.~(\ref{conservation})
remains valid even for $\Pi_{30}^{\alpha} \neq 0$. The only difference
is that the expressions (\ref{current-density.general})
for current and density
\begin{equation}
\label{current-density}
j_{\alpha} (i \Omega_n, \mathbf{K}) = \frac{e}{2}
(\mathbf{\bar{\Lambda}}^{\alpha \beta}(K) i K_{\beta} - i
 \mathbf{\bar{\Pi}}_{03}^{\alpha}(K) i \Omega_n) \theta(K), \qquad
\rho(i \Omega_n, \mathbf{K}) = - \frac{e}{2}
( \mathbf{\bar{\Pi}}_{33} (K)i \Omega_n +
i \mathbf{\bar{\Pi}}_{03}^{\alpha}(K) i K_{\alpha}) \theta(K)
\end{equation}
are now more complicated and contain  $\mathbf{\bar{\Pi}}_{03}^{\alpha}$.
Nevertheless substitution of  Eq.~(\ref{current-density})
in Eq.~(\ref{conservation}) indeed results in
Eq.~(\ref{phase.modes}) with $\mathcal{M}_{\theta}$ given by Eq.~(\ref{theta.cor.new}).
Hence, even in the most general case $\Pi_{03}^{\alpha} \ne 0$
the position of zeros of $M_{\theta}^{-1}$ (poles of $M_{\theta}$)
is gauge independent.

\subsection{Equation for the CG mode, its physical interpretation
and gauge independence}
\label{sec:CG.equation}

It can be checked that in the limit $\mathbf{K} \to 0$ and
$\Omega/K \to \infty$ one of the collective modes is the plasma
mode considered above. We are, however, more interested whether a
sound-like mode ($\Omega / K = \mbox{const}$ for $\mbox{K} \to 0$)
which would be similar to the BA mode in neutral superconductors
can exist in the charged system. It is seen from
Eq.~(\ref{theta.cor.new}) that if the ratio $\Omega/K$ is fixed
and one is interested in the low energy excitations with $\Omega
\to 0$, only the last term of Eq.~(\ref{theta.cor.new}) is
relevant because $\mathbf{{\bar \Pi}}_{33}(K)$ and $\mathbf{{\bar
\Pi}}_{03}^{\alpha}(K)$ are $\sim V_{c}^{-1}(\mathbf{K})$ in this
limit. In this case the equation
$\mathcal{M}_{\theta}^{-1}(\Omega, {\bf K})=0$ after the
analytical continuation $i \Omega_n \to \Omega + i0$ reduces to
\begin{equation}
\label{CG.equation}
\mathbf{\bar{\Lambda}}^{\alpha \beta}( \Omega, \mathbf{K}) K_{\alpha} K_{\beta} =
\left(\Lambda^{\alpha \beta}(\Omega, \mathbf{K})  -
\frac{\Pi_{03}^{\alpha}(\Omega, \mathbf{K})
\Pi_{30}^{\beta}(\Omega, \mathbf{K})}{\Pi_{33}(\Omega, \mathbf{K})} \right)
 K_{\alpha} K_{\beta} =0,
\end{equation}
so that the detailed form of the Coulomb interaction
becomes irrelevant, as was mentioned above. The solutions of
Eq.~(\ref{CG.equation}) are gauge independent
because for $\mathbf{K \to 0}$ this equation is equivalent
to the gauge independent equation
$\mathcal{M}_{\theta}^{-1}(\Omega, {\bf K})=0$.

It is interesting that practically the same arguments about the gauge independence
of the CG mode studied using the formalism of
Refs.~\cite{Kulik.1981,Takada.1988,Takada.1997,Takada.2000}
can be made. The corresponding equation for the CG mode is
\begin{equation}
\label{Takada.equation}
\mathbf{\bar \Pi}_{22}(\Omega, \mathbf{K}) - \Pi_{22}(0, \mathbf{0}) =0,
\end{equation}
where
\begin{equation}
\mathbf{\bar \Pi}_{22}(\Omega, \mathbf{K}) = \Pi_{22}(\Omega, \mathbf{K}) +
\Pi_{23}(\Omega, \mathbf{K})
\frac{V_c(\mathbf{K})}{1- V_{c}(\mathbf{K})  \Pi_{33}(\Omega, \mathbf{K}) }
\Pi_{32}(\Omega, \mathbf{K})
\end{equation}
$\Pi_{23}$ is given by Eq.~(\ref{Pi23}) and the definition of $\Pi_{22}$
is given in \cite{Takada.2000} (in addition to two obvious $\tau_2$ matrices
it also contains the factor $\gamma^2(\mathbf{k})$ because $d$-wave pairing is
considered).
One can show that in the limit $\mathbf{K} \to 0$  equation (\ref{Takada.equation})
reduces to the Ward identity
\begin{equation}
\label{indentity}
[-2/V + \Pi_{22}(\Omega, \mathbf{K} \to 0)] \Pi_{33}(\Omega, \mathbf{K} \to 0) +
\Pi_{23}^2(\Omega, \mathbf{K} \to 0) =0\,.
\end{equation}
Exactly this identity was recently proven in \cite{Smith.2000} (to be precise,
$s$-wave pairing was considered in \cite{Smith.2000})
using the gauge independence (charge conservation) arguments
(see e.g. \cite{Schrieffer}) and the mean field gap equation
\begin{equation}
\frac{2}{V} + \Pi_{22}(0, \mathbf{0}) =0.
\end{equation}
As was already mentioned the last equation has to be explicitly used in the
formalism of \cite{Kulik.1981,Takada.1988,Takada.1997,Takada.2000}.

It is possible to
establish a link between Eq.~(\ref{CG.equation}) and a simple and
transparent interpretation of the CG mode suggested by Schmid and
Sch\"on \cite{Schmid.1975} (see also Chap.~13 in \cite{book.1986}).
Comparing the expression for current in Eq.~(\ref{current-density})
and Eq.~(\ref{CG.equation}) one can notice that since
$\mathbf{{\bar \Pi}}_{03}^{\alpha}(K)$ disappears in the limit
$\mathbf{K} \to 0$, Eq.~(\ref{CG.equation}) is just a condition
\begin{equation}
\nabla \cdot \mathbf{j} = 0, \qquad
j_{\alpha} (i \Omega_n, \mathbf{K}) = \frac{e}{2}
\mathbf{\bar{\Lambda}}^{\alpha \beta}(K) i K_{\beta} \theta (K)
\end{equation}
which implies that if there is a supercurrent in the system, it
should be compensated by some normal current minimizing the total current.
Exactly the same condition which relates the CG mode to a counterflow
of supercurrent and normal current is discussed in \cite{book.1986}.
Thus using the two fluid picture one may say that
$\mathbf{\bar{\Lambda}}^{\alpha \beta}(K)$ in Eq.~(\ref{CG.equation}) consists of
the supercurrent $\Lambda^{\alpha \beta}$ and normal
$\mathbf{\bar{\Lambda}}^{\alpha \beta}(K) - \Lambda^{\alpha \beta}(K)$ parts,
respectively.

Although this counterflow resembles second sound in He$^{4}$, it
was pointed out in the earliest studies \cite{Schmid.1975} (see also
\cite{book.1980,book.1986}) that the CG mode is not the second
sound which is a hydrodynamic mode since in the CG mode the normal
fluid and superfluid  are not in the local thermodynamic
equilibrium and this mode is not a hydrodynamic mode.

It has to be stressed that even though the solutions of
Eqs.~(\ref{CG.equation}) and (\ref{Takada.equation}) are gauge
independent for $\mathbf{K} \to 0$, the existence of such
solutions {\em is not required} by the gauge invariance because no
general statement can be made about the polarizations $\Pi_{33}$,
$\Lambda^{\alpha \beta}$ and $\Pi_{03}^{\alpha}$ for arbitrary
values of $\Omega$ and $\mathbf{K}$. Thus, in contrast to the BA
and plasma modes  the existence of which is guaranteed by the
Goldstone theorem \cite{foot4} and Anderson-Higgs mechanism, the
CG mode does not obey any theorem and its existence is a fortunate
result of many subtle features of the system dynamics.

\subsection{Velocity of the CG mode}
\label{sec:velocity}

It is difficult to solve Eq.~(\ref{CG.equation}) analytically for
$d$-wave pairing and even numerically a more simple equation for
real part, $\mbox{Re} \mathbf{\bar{\Lambda}}^{\alpha
\beta}(\Omega, \mathbf{K}) =0$ is usually considered
\cite{Takada.1997,Takada.2000}. This significantly simplifies its
solution, but in general this approximation can be justified only
{\it a posteriory\/}, when the imaginary part is estimated. It is
possible to study this equation in two ways. The first way is to
extract the dispersion law $\Omega(\mathbf{K})$ for, in general,
arbitrary $\mathbf{K}$. The second way is to find the velocity of
the CG mode, $v_{CG}(\phi) \equiv \Omega / |\mathbf{K}|$ in the
limit $\mathbf{K} \to 0$. The extraction of the dispersion law is
more sensitive to the approximations which were made in the
calculation of the polarization operators. In particular, usage of
the approximated expressions (\ref{Pi_{33}}), (\ref{Pi_{00}}) and
(\ref{Pi_{03}}), which neglect the pair breaking for $\Omega \geq
2 \Delta_{0}$,  introduces the restriction $v_{F} K < \Delta_{0}$.
Moreover, for the polarizations Eqs.~(\ref{pi_{33}.definition}),
(\ref{pi_{00}}) and (\ref{pi_{03}}) calculated using the nodal
approximation (see Appendix~\ref{sec:B}) the condition of
smallness of $\mathbf{K}$ becomes even more strict, so that here
we study only the equation for $v_{CG}$. As also discussed in
Appendix~\ref{sec:B} the nodal approximation is valid for $T \ll
\Delta_{0}$. This is the reason why in the present paper only the
temperatures $T < 0.6 T_c$ are considered, where $T_c$ is defined
by Eq.~(\ref{Tc}). Nevertheless, the presented formalism  allows
in principle to study the phase fluctuation structure factor up to
$T_c$ if no additional approximations are made in calculation of
the polarizations.

\begin{figure}[h]
\centering{
\includegraphics[width=8cm]{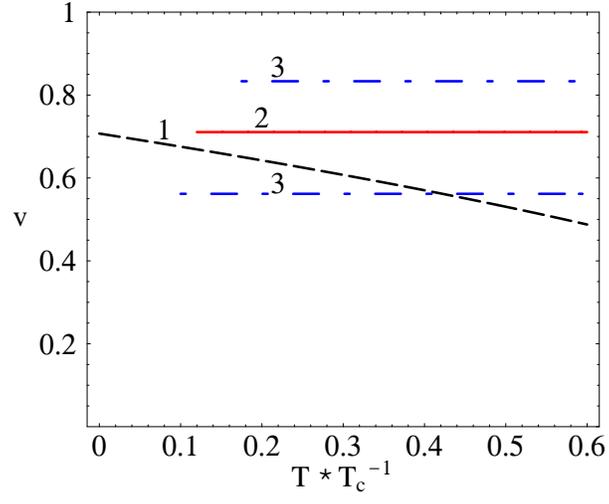}}
\caption{Temperature dependence of the velocities of the BA mode (dashed line -- $1$),
the CG mode for $\phi =0$ (full line -- $2$) and  $\phi =\pi/16$ (dot-dashed lines --
$3$). We put  $\alpha_{D} = 10$ and the velocities
are expressed in units of $v_{F}$.}
\label{fig:2}
\end{figure}

Thus for the illustrative purpose and comparison with \cite{Takada.2000}
we also solve numerically the equation
\begin{equation}\label{CG.final}
\mbox{Re} \mathbf{\bar{\Lambda}}^{\alpha \beta}
\left(\frac{\Omega}{K}= v_{CG}(\phi), K \to 0 \right)
\frac{K_{\alpha} K_{\beta}}{K^2} = 0
\end{equation}
(it is rewritten in dimensionless form (\ref{CG.dim}) for the numerical work)
instead of Eq.~(\ref{CG.equation}) and the results are presented in Fig.~\ref{fig:2}.
We stress that for $\phi = \pi/16$ we obtain two solutions:
$v_{CG} \approx 0.83 v_{F}$ and $v_{CG}\approx 0.56 v_{F}$.
These results are in fact in a very good agreement with the results shown in
Figs.~2 (a) and (b) of \cite{Takada.2000}. In particular, we also obtain that
the velocity of the CG mode is practically temperature independent and its
value is well described by $v_{CG} = v_{F} \cos(\phi - \pi/4)$
(or $v_{CG}^{\prime} = v_{F} \cos(\phi - \pi/4 + \pi/2)$
for the other pair of nodes which is rotated with respect to the first
pair by $\pi/2$). Indeed, this equation gives
\begin{equation}
\label{CG.position}
v_{CG} = v_{F} \times \left\{
\begin{array}{cccclc}
0.707 & (0.711 &\mbox{at}& T = 0.5 T_c) &: \phi = 0 & (\mbox{antinodal direction}) \\
0.831 & (0.833 &\mbox{at}&  T = 0.5 T_c )&: \phi = \pi/16 &   \\
0.555 &  (0.562 &\mbox{at}&  T = 0.5 T_c )&: \phi = \pi/16 &
(\mbox{other pair of nodes})\\
1     & (1.00 &\mbox{at}& T = 0.5 T_c) &: \phi = \pi/4 & (\mbox{nodal direction})
\end{array}
\right.
\end{equation}
Interestingly we obtain that the CG mode disappears even at
somewhat lower temperature than in \cite{Takada.2000}, $T = 0.1
T_c$. This lowest value of $T$ when the CG mode still exists is,
however, more the result of numerical solution of
Eq.~(\ref{CG.final}) than a real threshold  temperature, because
the peaks of the structure factor considered below disappear
rather gradually.

\subsection{Structure factor}
\label{sec:structure.factor}

The knowledge of the structure factor $S(\Omega, \mathbf{K})$ is
even more important than the value of the velocity, $v_{CG}$,
because this factor contains the information about the damping of
the CG mode \cite{foot5}. Furthermore, as we already mentioned,
this is the quantity which is measured in the Carlson-Goldman
experiment \cite{Goldman.discovery,Goldman.1981}.

One of the main results of the present paper is that we obtain a
closed {\em analytical} expression for the structure factor
(\ref{structure.factor.final}) for the clean $d$-wave
superconductor which is just a substitution of $\Pi_{33}$,
$\Pi_{03}^{\alpha}$ and $\Lambda$ given by
Eqs.~(\ref{pi_{33}.definition}), (\ref{pi_{03}}), (\ref{pi_{00}})
in Eqs.~(\ref{theta.cor.new}) and (\ref{theta.cor.new}) (its
inverse imaginary part) in Eq.~(\ref{structure.factor.final}),
respectively. The dimensionless form of $M_{\theta}^{-1}$ which
convenient for numerical evaluation is given in
Appendix~\ref{sec:D} (Eqs.~(\ref{M.theta.num}) and
(\ref{M.tilde})). In our formalism $M_{\theta}(\Omega,
\mathbf{K})$ for a fixed value of $v_{F} K$ depends on the
following dimensionless ratios: ($i$) $\alpha_{D} =
v_{F}/v_{\Delta}$, ($ii$) $T/\epsilon_{F} \equiv T/T_c (T_c
/\epsilon_{f})$ (see Eq.~(\ref{Tc}) where $T_c$ is expressed via
$\epsilon_{F}$), ($iii$) $v_F^2 K^2/ \omega_p^2 \equiv v_F^2
K^2/\epsilon_{F}^2 (\epsilon_{F}^2/\omega_p^{2})$ and the angle
$\phi$ which characterizes the direction of $\mathbf{K}$ with
respect to the Fermi surface (see Appendix~\ref{sec:D}). The ratio
$v_F^2 K^2/ \omega_p^2$ which was absent in Eq.~(\ref{CG.final})
is now present because the full expression for $M_{\theta}^{-1}$
contains the Coulomb potential $V_{c}(\mathbf{K})$ (see its
representation in terms of $v_{F} K$ and $\omega_p$ in
Eq.~\ref{Coulomb}) which we chose in the simplest 3D form. It is
indeed easy to see that since $v_{F} K \ll \omega_{p}$, the
detailed form of $V_{c}$ is not important and here we compute
$S(\Omega, \mathbf{K})$ including the Coulomb potential only to
demonstrate this explicitly.

Despite the gauge dependent definition of the pair susceptibility
Eq.~(\ref{pair.susceptibility}), the peaks of the structure factor
Eq.~(\ref{structure.factor.final}) considered below which are associated with
the singularities of $\mathcal{M}_{\theta}(\Omega, \mathbf{K})$ (zeros of
$\mathcal{M}_{\theta}^{-1}(\Omega, \mathbf{K})$) can be regarded as
{\em gauge independent} in the sense that the position of these singularities
is gauge independent as was argued above.  
Furthermore, as shown in \cite{Gusynin} a more general 
expression for $\mathcal{M}_{\theta}$ derived from Eq.~(\ref{V.P.})
{\it does depend} on the gauge fixing parameter $\lambda$, but in a such way
that the position of the {\it pole} of $\mathcal{M}_{\theta}$ and 
{\it its residue} are {\it gauge invariant}. This justifies the use
of the structure factor $S(\Omega, \mathbf{K})$ which is expressed via
$\mbox{Im} \mathcal{M}_{\theta}(\Omega, \mathbf{K})$. 

\begin{figure}[h]
\centering{
\includegraphics[width=8cm]{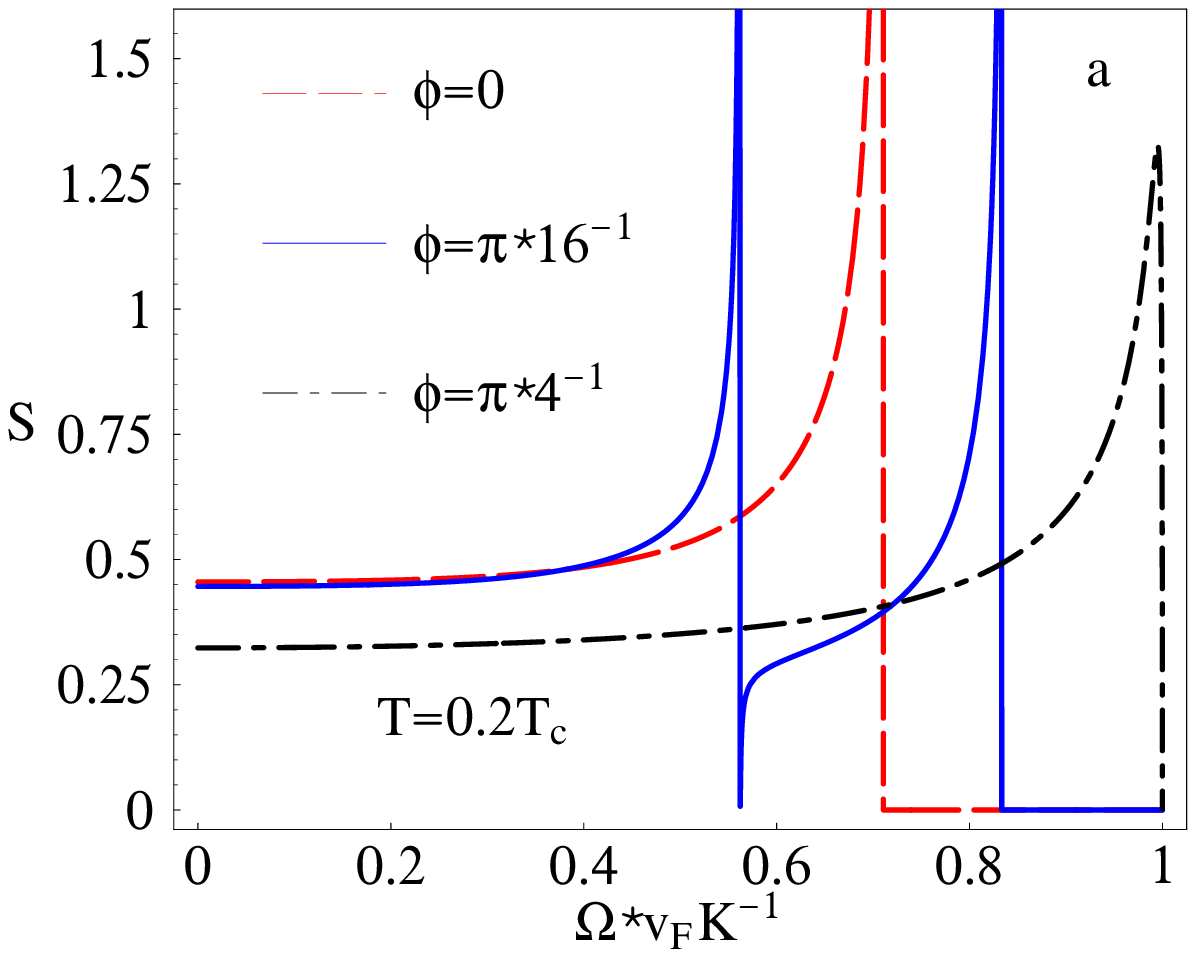} \\
\includegraphics[width=8cm]{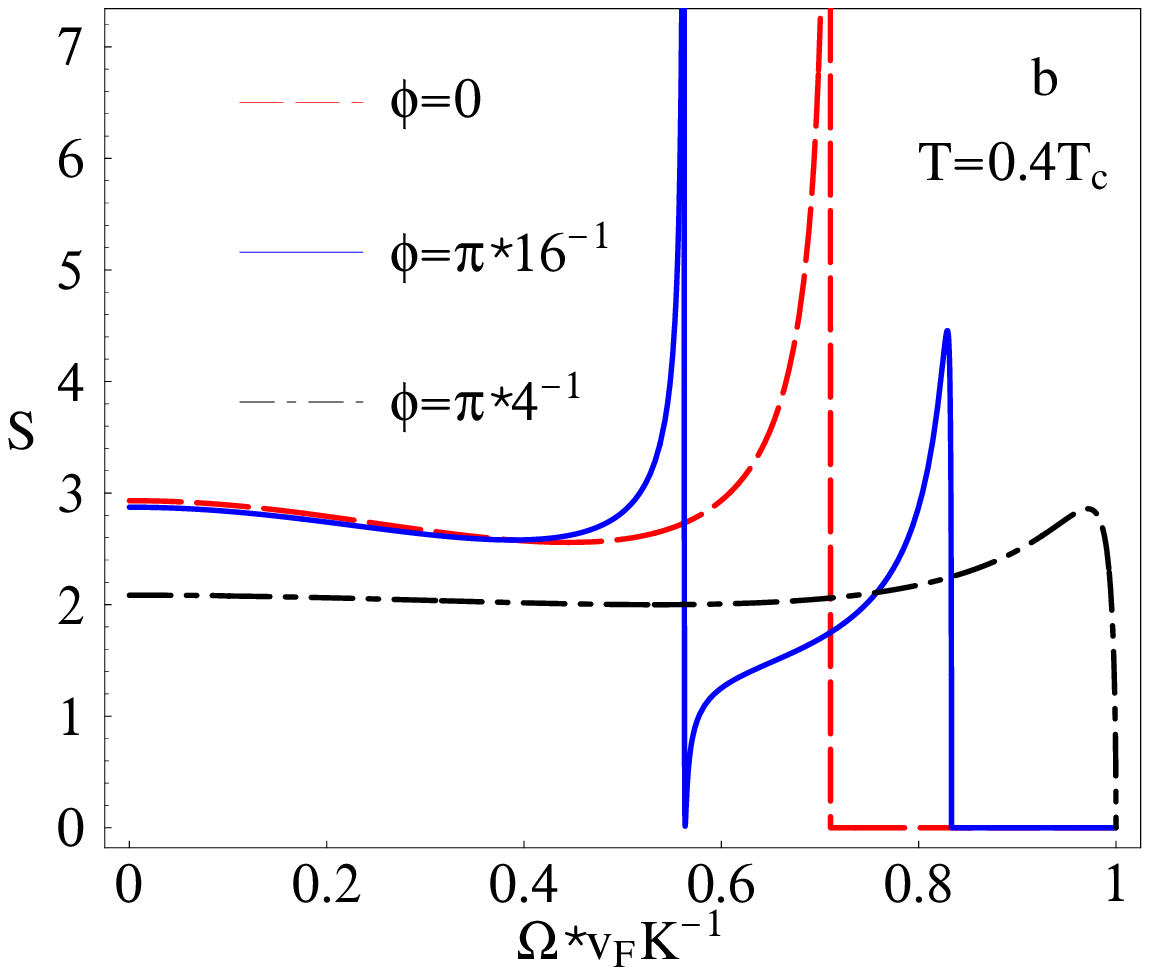} \\
\includegraphics[width=8cm]{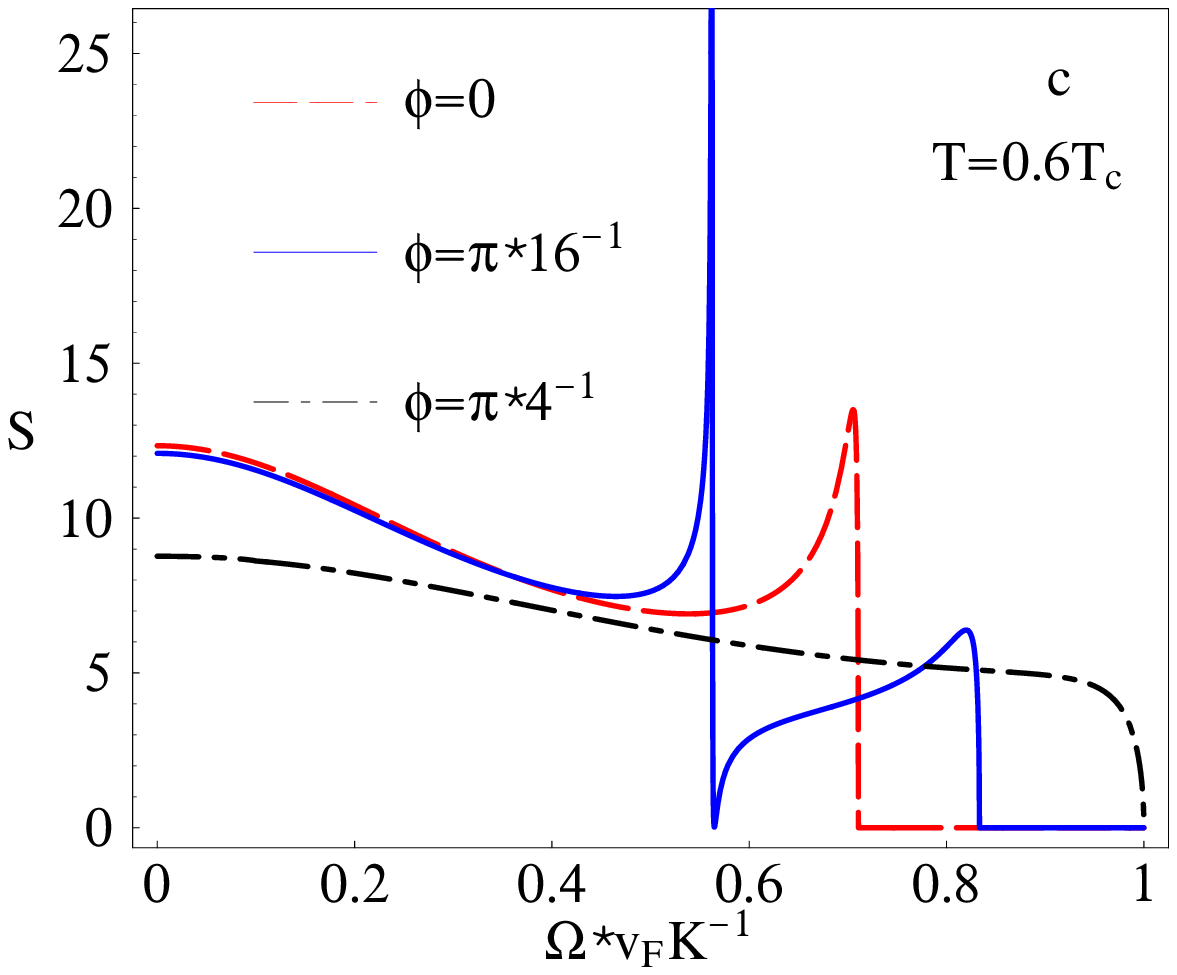}}

\caption{Structure factor $S(\Omega, \mathbf{K})$ in arbitrary units:
(a)  $T = 0.2 T_c$, (b) $T = 0.4 T_c$ and (c) $T = 0.6 T_c$.
We put $\omega_p = 0.5 \epsilon_{F}$, $v_{F} K = 0.01 \epsilon_{F}$
(i.e. $v_{F} K = 0.02 \omega_p$) and $\alpha_{D} = 10$, so that
$T_{c} \approx  0.063 \epsilon_{F}$.}
\label{fig:3}
\end{figure}

In Fig.~\ref{fig:3} (a) - (c)
we show the structure factor calculated using
the analytical expressions mentioned above for different temperatures.
The position of the peaks defined by the ratio $\Omega/v_{F} K$ is well
fitted by Eq.~(\ref{CG.position}). In particular, for $\phi = \pi/16$ two peaks
are seen. This allows to associate the origin of these peaks with the gap nodes.
As expected the peaks are getting less sharp and higher as the temperature
increases. The width of the peaks also depends on the direction of $\mathbf{K}$:
for a larger value of $\phi$ ($0 < \phi < \pi/4$) the corresponding peak
is wider. All these results are in agreement with Fig.~3 from \cite{Takada.2000}, but
due to the analytical character of the calculation the subtle peak features
have a better resolution.

\begin{figure}[h]
\centering{
\includegraphics[width=8cm]{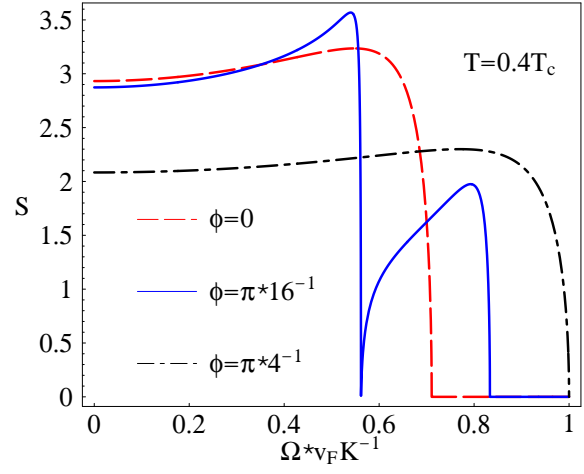}}

\caption{Structure factor $S(\Omega, \mathbf{K})$ in arbitrary
units for $T = 0.4 T_c$. The rest of the parameters is the same as
in Fig.~\ref{fig:3}, except for the fact that the density-current
coupling is ``switched off'' ($\Pi_{03}^{\alpha} =0$.)}
\label{fig:4}
\end{figure}

\begin{figure}[h]
\centering{
\includegraphics[width=8cm]{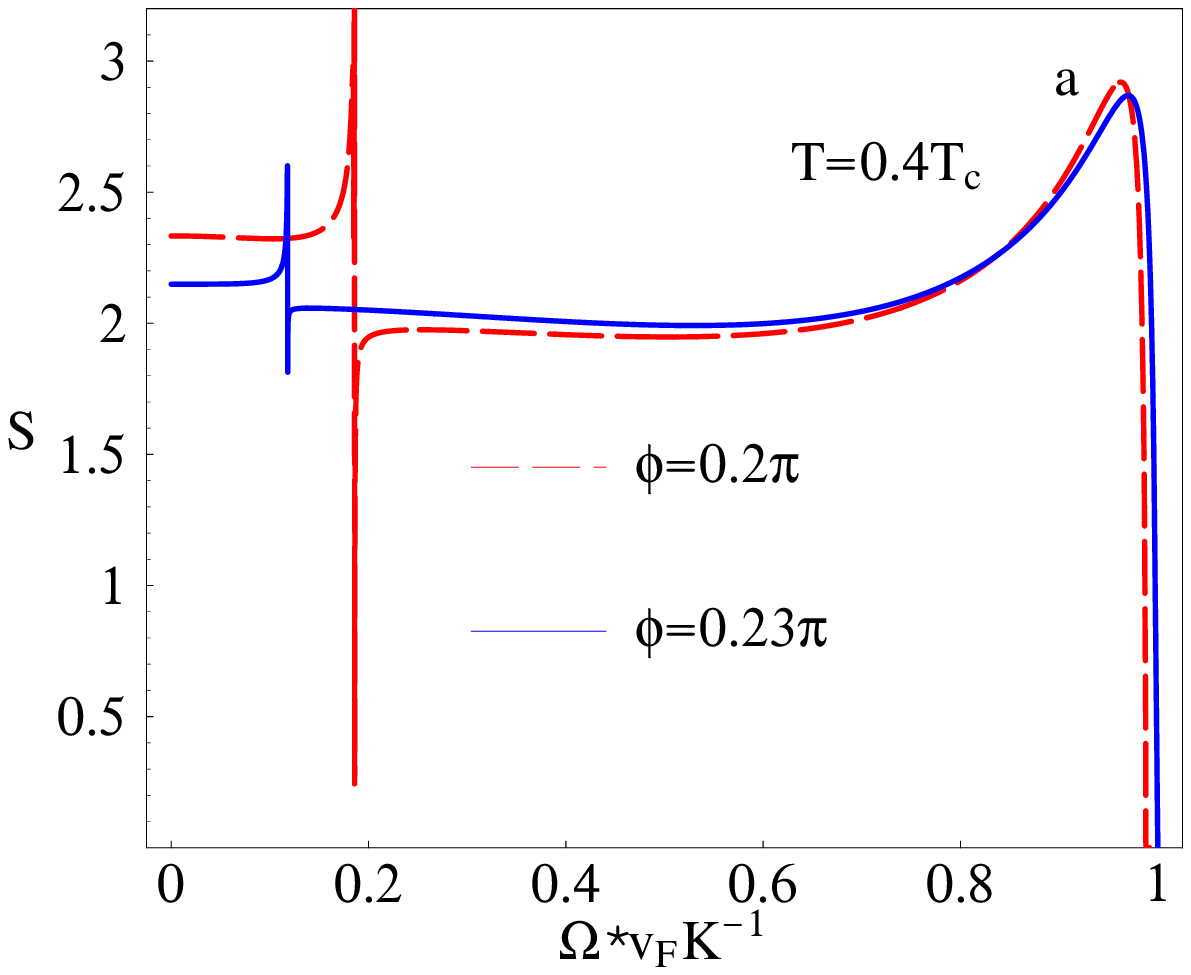} \\
\includegraphics[width=8cm]{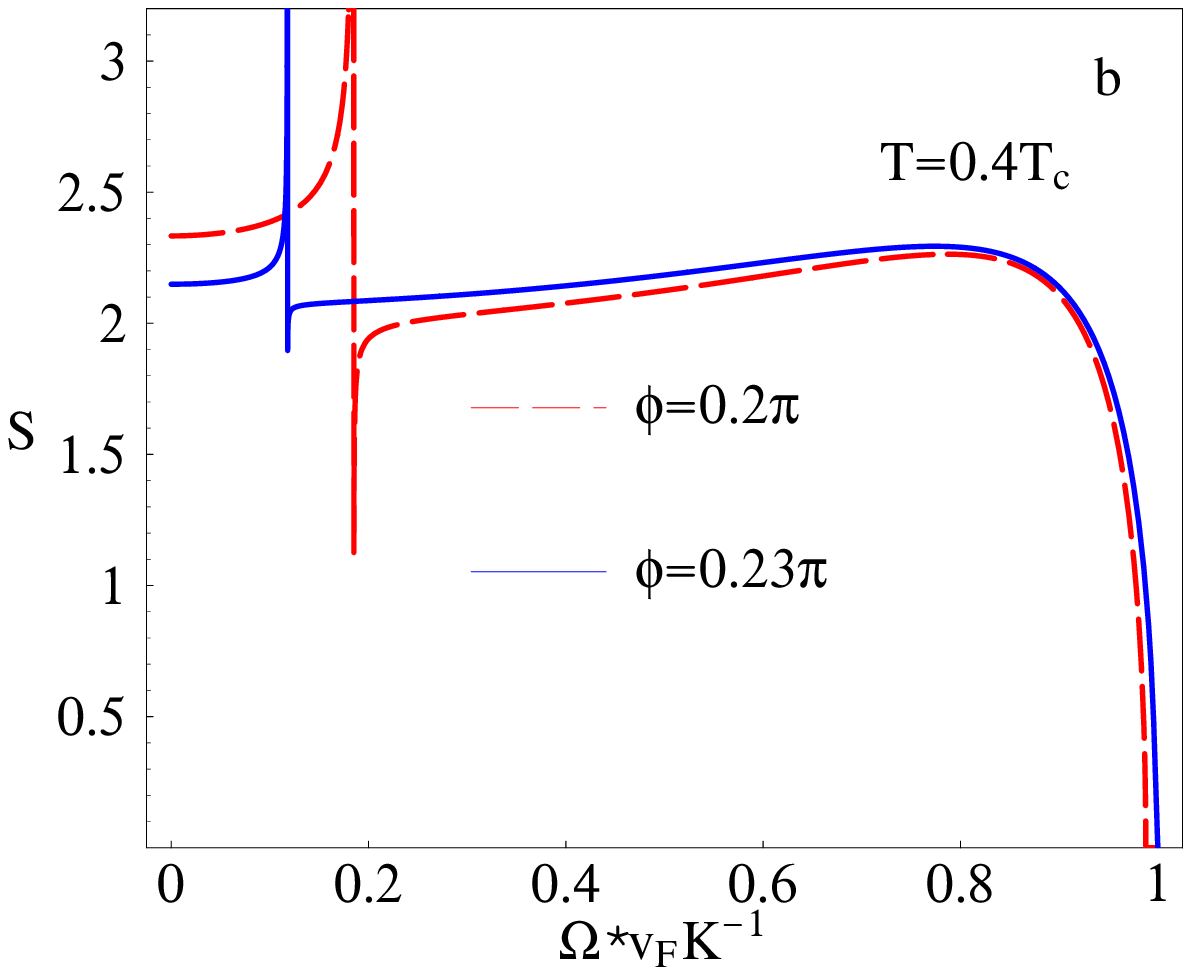}}

\caption{Structure factor $S(\Omega, \mathbf{K})$ in arbitrary units for
$T = 0.4 T_c$.
(a)  $\Pi_{03} \neq 0$, (b) $\Pi_{03} = 0$.
The rest of the parameters is the same as in Fig.~\ref{fig:3}.}
\label{fig:5}
\end{figure}

To argue that these peaks in Fig.~\ref{fig:3} are indeed due to the density-current
coupling in Fig.~\ref{fig:4} we show for comparison the structure factor which
was calculated setting $\Pi_{30}^{\alpha} =0$. The disappearance of the peaks
confirms our claim that the CG mode demands nonzero density-current coupling.

This procedure of setting $\Pi_{03}^{\alpha}$ equal to zero is in fact very convenient
in clarifying the origin of the peaks, because even for $\Pi_{03}^{\alpha}$
some peaks can be seen because both $\Pi_{33}$ and $\Pi_{00}^{\alpha \beta}$
have the square root singularities discussed in Sec.~\ref{sec:Omega}.
For example, for $\pi/16 < \phi < \pi/4$ (see Fig.~\ref{fig:5}) the lower peak
becomes even sharper, but in contrast to to Fig.~\ref{fig:4} it
{\em does not disappear} when we set $\Pi_{03}^{\alpha} =0$.

\subsection{Manifestations of the CG mode in the classical response functions}
\label{sec:correlators}

The structure factor considered above is so important for the
investigation of the CG mode because for superconductors there is
no a classical laboratory field that couples, to and in the static
limit is the thermodynamic conjugate of the order parameter. The
reason that there is no laboratory conjugate field in the
superconductor and superfluid cases is that these order parameters
are off-diagonal in number space \cite{Goldman.1981}. Nevertheless
it is interesting to investigate whether the CG mode can manifest
itself in the ``classical'' correlators when one goes beyond the
static limit.

Let us firstly consider the form $M_{\varphi}$ for the electric potential.
Substituting the elements of the matrix Eq.~(\ref{M}) in Eq.~(\ref{varphi})
we arrive at the following expression (compare with Eq.~(\ref{varphi.standard}))
\begin{equation}
\begin{split}
\mathcal{M}_{\varphi}^{-1}(i\Omega_n, {\bf K})   = 4 e^2
\left(
\frac{- \Pi_{33} \Lambda^{\alpha \beta} K_{\alpha} K_{\beta}
+ \Pi_{03}^{\alpha} \Pi_{30}^{\beta} K_{\alpha} K_{\beta}}
{-\Omega_n^2 \Pi_{33} + \Lambda^{\alpha \beta} K_{\alpha} K_{\beta}
- i \Omega_n K_{\alpha} \Pi_{03}^{\alpha} -
i \Omega_n K_{\alpha} \Pi_{30}^{\alpha}} + V_{c}^{-1}(\mathbf{K})
\right).
\end{split}
\end{equation}
One can verify that both plasma and CG modes are present in the equation
$\mathcal{M}_{\varphi}^{-1} = 0$.

Although the equation $\mathcal{M}_{\varphi}^{-1} = 0$ is gauge
independent, $\mathcal{M}_{\varphi}^{-1}$ itself as well as
$\mathcal{M}_{\theta}^{-1}$ is gauge dependent. This time,
however, we can take instead of $\mathcal{M}_{\varphi}$ truly
gauge independent correlators Eq.~(\ref{cor.def}). Note that the
definition Eq.~(\ref{cor.def}) with the source fields $(A_0,
\mathbf{A})$ clearly shows that the ``classical laboratory field''
is used to probe the corresponding response. Then repeating the
calculation of Eqs.~(\ref{density.cor.standard}) and
(\ref{current.cor.standard}) with nonzero $\Pi_{03}^{\alpha}$ we
arrive at the gauge independent density-density
\begin{equation}
\label{density.cor.new}
\chi(i \Omega_n, {\bf K}) =
\frac{e^2 (\mathbf{\bar{\Pi}}_{03}^{\alpha} \mathbf{\bar{\Pi}}_{30}^{\beta}
- \mathbf{\bar{\Pi}}_{33} \mathbf{\bar{\Lambda}}^{\alpha \beta})
K_{\alpha} K_{\beta}}
{ \mathbf{\bar{\Lambda}}^{\alpha \beta} K_{\alpha} K_{\beta} -
\mathbf{\bar{\Pi}}_{33} \Omega_n^{2}
- i \Omega_n K_{\alpha} \mathbf{\bar{\Pi}}_{03}^{\alpha}
- i \Omega_n K_{\alpha} \mathbf{\bar{\Pi}}_{30}^{\alpha}}.
\end{equation}
and current-current
\begin{equation}
\label{current.cor.new}
\chi^{\alpha \beta}(i \Omega_n, {\bf K}) = e^2 \left(
\mathbf{\bar{\Lambda}}^{\alpha \beta}
- \frac{(i \Omega_n \mathbf{\bar{\Pi}}_{03}^{\alpha} -
\mathbf{\bar{\Lambda}}^{\alpha \mu} K_{\mu})
(i \Omega_n \mathbf{\bar{\Pi}}_{30}^{\beta} -
\mathbf{\bar{\Lambda}}^{\nu \beta} K_{\nu})}
{ \mathbf{\bar{\Lambda}}^{\alpha \beta} K_{\alpha} K_{\beta} -
\mathbf{\bar{\Pi}}_{33} \Omega_n^{2}
- i \Omega_n K_{\alpha} \mathbf{\bar{\Pi}}_{03}^{\alpha}
- i \Omega_n K_{\alpha} \mathbf{\bar{\Pi}}_{30}^{\alpha}} \right),
\end{equation}
correlators.

\begin{figure}[h]
\centering{
\includegraphics[width=8cm]{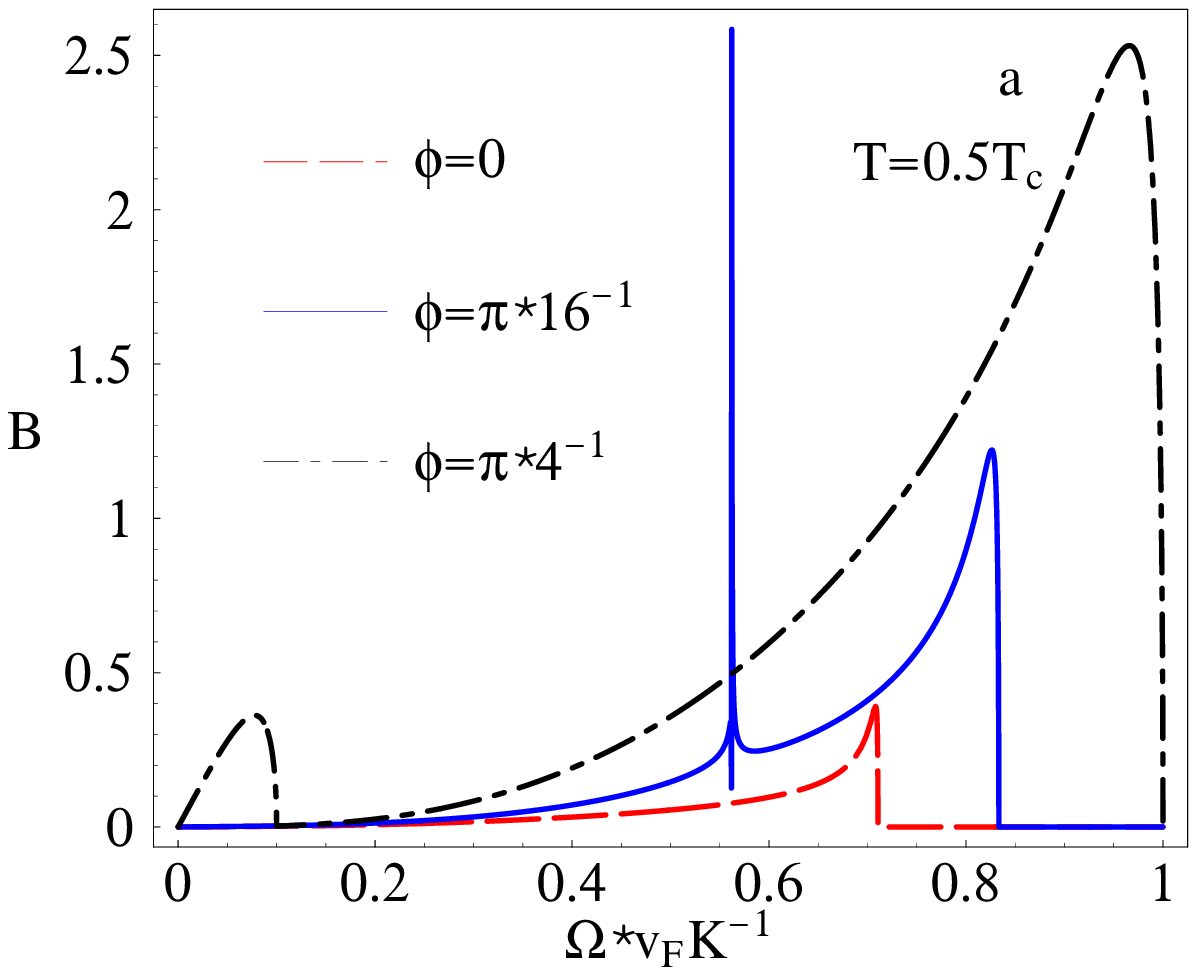}\\
\includegraphics[width=8cm]{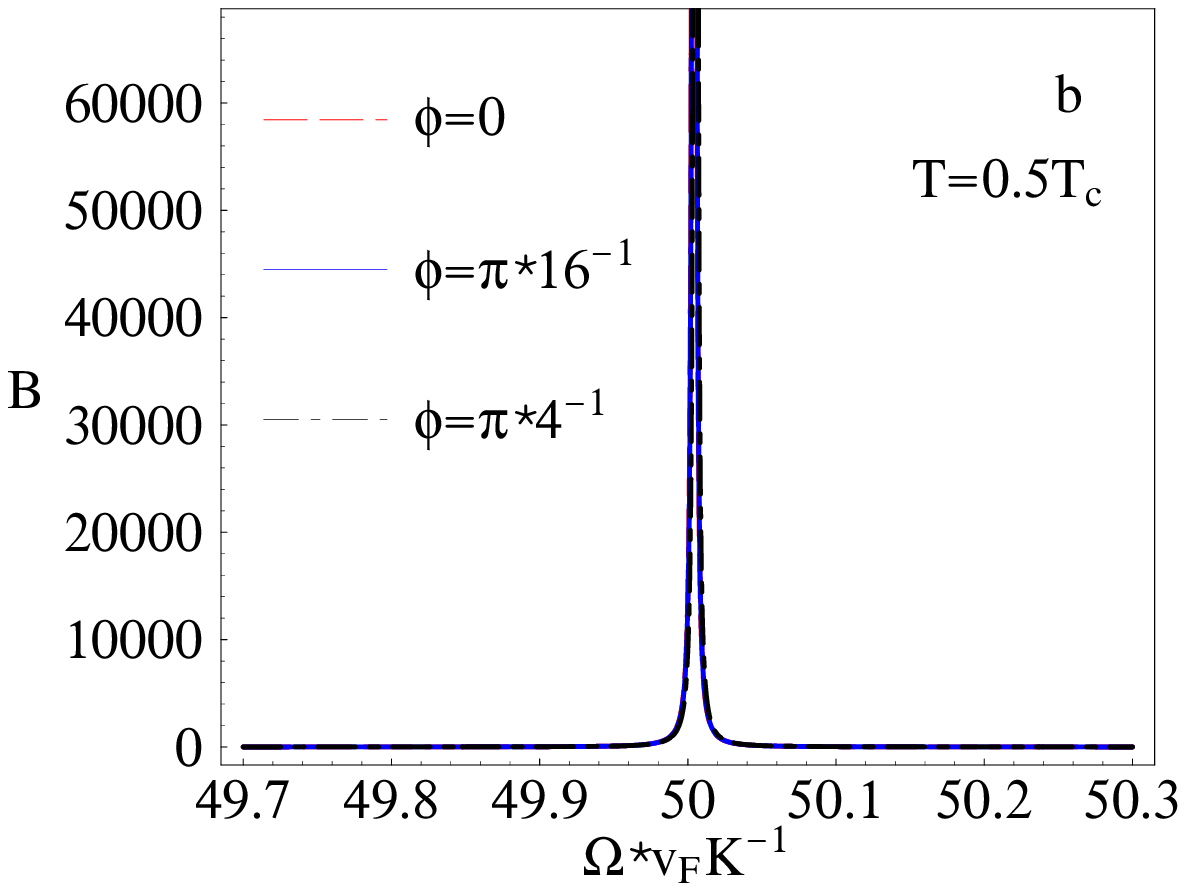}\\
\includegraphics[width=8cm]{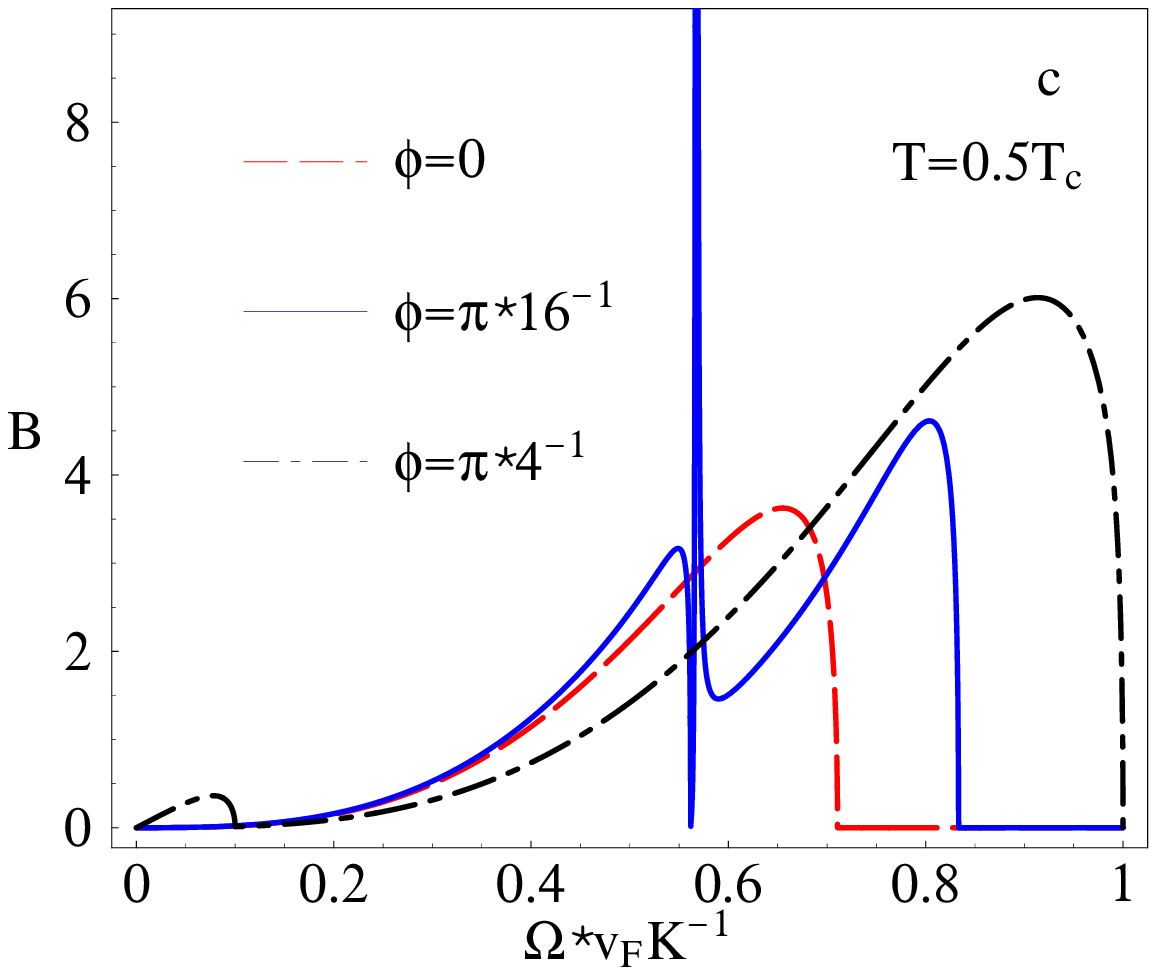}}

\caption{Spectral density $B(\Omega, \mathbf{K})$ in arbitrary units
for $T = 0.5 T_c$.
(a) $0 \leq \Omega/v_{F} K \leq 1$ and $\Pi_{03}^{\alpha} \ne 0$,
(b) $49 \leq \Omega/v_{F} K \leq 51$ and $\Pi_{03}^{\alpha} \ne 0$,
(c) $0 \leq \Omega/v_{F} K \leq 1$ and $\Pi_{03}^{\alpha} = 0$.
We put $\omega_p = 0.5 \epsilon_{F}$, $v_{F} K = 0.01 \epsilon_{F}$
(i.e. $ \omega_p = 50 v_{F} K$) and $\alpha_{D} = 10$.
To display the $\delta$-like plasma peak seen in Fig.~\ref{fig:6} (b) a small
imaginary part is added to $\omega_p$.}
\label{fig:6}
\end{figure}

In Fig.~\ref{fig:6} we show the spectral density
\begin{equation}
B(\Omega, \mathbf{K}) = \frac{1}{\pi} \mbox{Im} \chi(\Omega, \mathbf{K})
\end{equation}
calculated for the density-density correlator Eq.~(\ref{density.cor.new}).
The plasma mode (see Fig.~\ref{fig:6} (b)) and the CG mode (see Fig.~\ref{fig:6} (a))
as proves Fig.~\ref{fig:6} (c) are clearly seen in the density-density
correlator. We note that  for $\phi = \pi/16$ the lower peak does not change strongly
showing that it is due to the density-density correlations and not the CG mode.

\begin{figure}[h]
\centering{
\includegraphics[width=8cm]{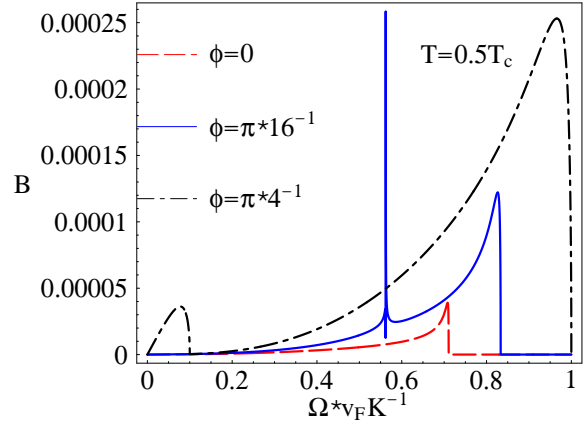}}

\caption{Spectral density $B(\Omega, \mathbf{K})$ in arbitrary but the same
as in Fig.~\ref{fig:6} units for $T = 0.5 T_c$.
We put $\omega_p = 0.5 \epsilon_{F}$, $v_{F} K = 0.001 \epsilon_{F}$
(i.e. $ \omega_p = 500 v_{F} K$) and $\alpha_{D} = 10$.}
\label{fig:7}
\end{figure}

As the value of $K$ decreases the relative weight of the CG
mode with respect to the plasma mode becomes smaller and smaller
as shown in Fig.~\ref{fig:7}.  Finally in the limit
\begin{equation}
\chi(i \Omega_n, {\bf K}) \sim -
\frac{e^2 \mathbf{\bar{\Pi}}_{33} \mathbf{\bar{\Lambda}}^{\alpha \beta}
K_{\alpha} K_{\beta}}
{\mathbf{\bar{\Lambda}}^{\alpha \beta} K_{\alpha} K_{\beta}},
\qquad \frac{\Omega}{K} = \mbox{const}, \mathbf{K} \to 0,
\end{equation}
the CG mode disappears from the density-density correlator.
This agrees with the statement of \cite{Takada.1998} that the density-density
correlator in the limit $\mathbf{K} \to 0$ is dominated by the plasma oscillation
only.

Treating the current-current correlator Eq.~(\ref{current.cor.new}) in the same way
one obtains
\begin{equation}
\chi^{\alpha \beta}(i \Omega_n, {\bf K}) \sim e^2 \left(
\mathbf{\bar{\Lambda}}^{\alpha \beta}
- \frac{
\mathbf{\bar{\Lambda}}^{\alpha \mu} K_{\mu}
\mathbf{\bar{\Lambda}}^{\nu \beta} K_{\nu}}
{ \mathbf{\bar{\Lambda}}^{\alpha \beta} K_{\alpha} K_{\beta}} \right),
\qquad \frac{\Omega}{K} = \mbox{const}, \Omega \to 0.
\end{equation}
Taking into account that the structure of
$\mathbf{\bar{\Lambda}}^{\alpha \beta}$ is
$\mathbf{\bar{\Lambda}}^{\alpha \beta} = \Lambda \delta_{\alpha
\beta} - F v_{F \alpha} v_{F \beta}$ one can check that the term
with $F$ cancels out from the transverse correlator. This shows
that the longitudinal CG mode cannot be seen in the transverse
current-current correlator and its presence does not have any
influence on the Meissner effect in the limit $\mathbf{K} \to 0$.

\section{Discussion}
\label{sec:discussion}

In the present paper we have shown that the 2D model of clean
$d$-wave superconductor predicts the existence of the CG mode in a
wide temperature region down to $T = 0.1 T_c$. This is done using
the analytical expression for the structure factor which has peaks
associated with the CG mode and solving numerically the equation
for the CG mode velocity, $v_{CG}$. All our results are in a good
agreement with the paper \cite{Takada.2000} where a similar model
has been studied numerically using the formalism of
\cite{Kulik.1981}. It was also shown in \cite{Takada.2000} that in
contrast to $s$-wave superconductors where the impurities
supressing the Landau damping result in more favourable conditions
for the observation of the CG mode, in $d$-wave superconductors
the CG mode disappears in the dirty system. 

Thus the main physical question is whether our prediction of the
CG mode in a clean $d$-wave superconductor is relevant for HTSC
cuprates which are very complex compounds. Recent measurements
\cite{Zhang} done in high-purity YBa$_2$Cu$_3$O$_7$ crystals show
that the in-plane mean free path $l$ increases to $\simeq 1 \mu m$
below $T = 20 K (\approx 0.22 T_c)$. This suggests that these
systems are deeply in the clean limit $l \gg \xi \sim 10 \dot{A}$
($\xi$ is the in-plane superconducting coherence length) and the
model we considered can be applied to describe the CG mode in
these compounds. Using the value of the Fermi velocity $v_{F}
\approx 1.8 \times 10^7 cm/s$ from \cite{Zhang} we predict that
the velocity of the CG mode is expected to be within the range
$v_{CG} = (1/\sqrt{2} \, -\, 1) v_{F} \approx (1.3 \, -\, 1.8)
\times 10^5 m/c$ depending on the direction of $\mathbf{K}$ which
is one or two orders of magnitude faster than the velocity of the
CG mode observed in conventional superconductors.

From the theoretical point of view there are many questions which
deserve further theoretical investigation. First of all, it would
still be important to consider the influence of impurities and
inelastic scattering by antiferromagnetic spin fluctuations within
the proposed formalism. It is also interesting to estimate the
widening of the structure factor peaks which is discussed above
Eq.~(\ref{peak}). This widening may become important if
temperatures $T \lesssim T_c$ are considered.

The discovery of the CG mode in conventional superconductors led to much deeper
understanding of superconductivity, so we hope that the investigation of the same problem
in HTSC would also increase understanding of these complex systems.

\section{Acknowledgments}

We gratefully acknowledge Prof. V.P.~Gusynin for numerous fruitful discussions,
careful reading of the manuscript and his patience in explaining us 
the issue of gauge invariance.
We also thank Prof.~A.~Fetter for useful discussion and Prof.~V.M.~Loktev for
critical remarks on the manuscript and Prof.~P.Martinoli for bringing
our attention to the existence of the CG mode and his constant
interest to our progress.
S.G.Sh. is  grateful to the members of the Institut de Physique,
Universit\'e de Neuch\^atel for hospitality.
This work was supported by the research project 2000-061901.00/1
of the Swiss National Science Foundation.

\appendix
\section{General expressions for polarizations $\Pi_{ij}$}
\label{sec:A}

The general expressions for the polarization functions
$\Pi_{ij}$ are (see e.g. \cite{Takada.1997,we.LD,Aitchison.2000})
\begin{equation}
\label{Pi.general}
\begin{split}
& \begin{bmatrix}
\Pi_{00}^{\alpha \beta}(i \Omega_{n}, {\bf K}) \\
\Pi_{33}(i \Omega_{n}, {\bf K})
\end{bmatrix} = \\
\\ &  - \int \frac{d^2 k}{(2 \pi)^2}  \left\{ \frac{1}{2} \left(1 -
\frac{\xi_{-}\xi_{+} \pm \Delta_{-} \Delta_{+} }{E_{-}E_{+}} \right) \left[
\frac{1}{E_{+} + E_{-} + i \Omega_n} +
\frac{1}{E_{+} + E_{-} - i \Omega_n}\right] [1 - n_F(E_{-}) - n_F(E_{+})]
\right. \\
& \left. + \frac{1}{2} \left(1 + \frac{\xi_{-}\xi_{+} \pm \Delta_{-} \Delta_{+}
}{E_{-}E_{+}} \right)\left[ \frac{1}{E_{+} - E_{-} + i \Omega_n} +
\frac{1}{E_{+} - E_{-} - i \Omega_n}\right] [n_F(E_{-}) - n_F(E_{+})]
\right\}V_{\pm}(\mathbf{k})\,,\\
& \qquad \qquad \qquad V_{\pm}(\mathbf{k}) \equiv
\begin{bmatrix}
v_{F \alpha}(\mathbf{k}) v_{F \beta}(\mathbf{k}), & ``+''; \\
1, & ``-''.
\end{bmatrix}\,,
\end{split}
\end{equation}
and
\begin{equation}\label{Pi_{03}.general}
\begin{split}
& \Pi_{03}^{\alpha}(i \Omega_{n}, {\bf K}) =\\
& =  \int \frac{d^2 k}{(2 \pi)^2} \left\{ \left(\frac{\xi_{+}}{2E_{+}} -
\frac{\xi_{-}}{2E_{-}} \right) \left[ \frac{1}{E_{+} + E_{-} + i \Omega_n} -
\frac{1}{E_{+} + E_{-} - i \Omega_n}\right] [1 - n_F(E_{-}) - n_F(E_{+})]
\right. \\
& \left. + \left(\frac{\xi_{+}}{2E_{+}} + \frac{\xi_{-}}{2E_{-}} \right) \left[
\frac{1}{E_{+} - E_{-} + i \Omega_n} - \frac{1}{E_{+} - E_{-} - i
\Omega_n}\right] [n_F(E_{-}) - n_F(E_{+})] \right\}
v_{F \alpha}(\mathbf{k}),
\end{split}
\end{equation}
where $\xi_{\pm} \equiv \xi({\bf k} \pm {\bf K}/2)$, $E_{\pm} \equiv E({\bf k}
\pm {\bf K}/2)$ and $\Delta_{\pm} \equiv \Delta({\bf k} \pm {\bf K}/2)$.
One can also check that  $\Pi_{30}^{\alpha}(i\Omega_{n},{\bf K}) =
\Pi_{03}^{\alpha} (i\Omega_{n},{\bf K})$.

The first and second terms in Eqs.~(\ref{Pi.general}) and
(\ref{Pi_{03}.general}) have a clear physical interpretation
\cite{Schrieffer}. The first term proportional $1 - n_{F}(E_{-}) -
n_{F}(E_{+})$ gives the contribution from ``superfluid''
electrons. The second term gives the contribution of the thermally
excited quasiparticles (i.e. ``normal'' fluid component). This is
the term responsible for the appearance of the {\em Landau terms}
in the effective action (see e.g. \cite{Aitchison.2000,we.LD}).
The imaginary part of these terms is the only source of  damping
of the phase excitations in the clean system considered here. The
physical origin of this damping is due to the scattering of the
thermally excited quasiparticles from the phase excitations.

Since in what follows we are interested in the limit
$\Omega, v_{F} K \ll \Delta_{0}$ we may safely neglect $\Omega$
and $K$ in the first terms of  Eqs.~(\ref{Pi.general}),
(\ref{Pi_{03}.general}) and write
\begin{equation}\label{Pi_{33}}
\Pi_{33}(\Omega, {\bf K}) \approx - \int \frac{d^2 k}{(2 \pi)^2} \left\{
\frac{\Delta^2({\bf k})}{E^3({\bf k})} \tanh \frac{E({\bf k})}{2T}
+ \frac{\xi^2({\bf k})}{E^2({\bf k})}
\frac{2(E_{+} - E_{-})}{(E_{+} - E_{-})^2 - \Omega^2}
[n_{F}(E_{-}) - n_{F}(E_{+})] \right\},
\end{equation}
\begin{equation}\label{Pi_{00}}
\Pi_{00}^{\alpha \beta}(\Omega, {\bf K}) \approx - \int \frac{d^2 k}{(2 \pi)^2}
\frac{2(E_{+} - E_{-})}{(E_{+} - E_{-})^2 - \Omega^2}
[n_{F}(E_{-}) - n_{F}(E_{+})]
v_{F \alpha}({\bf k}) v_{F \beta}({\bf k}),
\end{equation}
\begin{equation}\label{Pi_{03}}
\Pi_{03}^{\alpha}(\Omega, {\bf K}) \approx  - \int \frac{d^2 k}{(2 \pi)^2}
\frac{\xi({\bf k})}{E({\bf k})}
\frac{2 \Omega}{(E_{+} - E_{-})^{2} - \Omega^2} [n_{F}(E_{-}) - n_{F}(E_{+})]
v_{F \alpha}({\bf k}).
\end{equation}
This approximation is in agreement with the one used in Appendix~C
of \cite{Takada.1997}.

\section{Nodal approximation}
\label{sec:B}

The key values which are necessary for evaluation of the polarizations
Eqs.~(\ref{Pi_{33}}), (\ref{Pi_{00}}) and (\ref{Pi_{03}})
are the differences $E_{+} - E_{-}$ and $n_{F}(E_{+}) -n_{F}(E_{-})$.
Expanding in $\mathbf{K}$,
\begin{equation}\label{difference.defintion}
E({\bf k} + {\bf K}/2) - E({\bf k} - {\bf K}/2) =
\mathbf{v}_{g}(\mathbf{k}) \mathbf{K}\,,
\end{equation}
where the quasiparticle group velocity is given by
\begin{equation}\label{group.velocity}
\mathbf{v}_{g}(\mathbf{k}) = \nabla_{\mathbf{k}} E(\mathbf{k}) =
\frac{1}{E(\mathbf{k})} \left[\xi(\mathbf{k}) \mathbf{v}_{F} +
\Delta(\mathbf{k}) \mathbf{v}_{\Delta}\right]\,.
\end{equation}
It is obvious that due to the gap $\mathbf{k}$-dependence
Eq.~(\ref{difference.defintion}) differs from the $s$-wave case
\cite{Aitchison.2000} by the second term \cite{Lee}.
To perform the calculation analytically
it is useful to rewrite Eqs.~(\ref{difference.defintion}) and
(\ref{group.velocity}) in terms of the nodal approximation described
in detail in \cite{Lee} (see also Sec.~VII of \cite{we.LD} where
the imaginary parts of these polarizations were calculated.)
In particular, using this approximation one has
$\varepsilon(\mathbf{k}) \simeq p_1 \equiv v_{F} k_1 = p \cos \delta$,
$\Delta(\mathbf{k}) \simeq p_2 \equiv  v_{\Delta} k_2 = p \sin \delta$
with  $E(\mathbf{k}) \simeq p =
\sqrt{p_1^2 + p_2^2} = \sqrt{v_{F}^2 k_1^2 + v_\Delta^2 k_2^2}$,
where the quasiparticle momentum $\mathbf{k} = (k_1, k_2)$ is
written in the nodal coordinate system
$\hat{\mathbf{k}}_{1}$, $\hat{\mathbf{k}}_{2}$
associated with $j$-th node ($j =1,\ldots,4$). (Note that the angle $\delta$
was denoted in \cite{Lee, we.LD} as $\varphi$).
Then
\begin{equation}\label{difference}
{\bf v}_{g} {\bf K} = v_{F} K_1 \cos \delta +
v_\Delta K_2 \sin \delta \equiv P \cos(\delta - \psi)\,,
\qquad E({\bf k} \pm {\bf K}/2) \ll \Delta_{0}\,,
\end{equation}
where the momentum $\mathbf{K} = (K_1,K_2)$ of $\theta$-particle is also
expressed in the nodal coordinate system
$\hat{\mathbf{k}}_{1}$, $\hat{\mathbf{k}}_{2}$, so that
$P^1 \equiv v_{F} K_1  = P \cos \psi$, $P^2 \equiv v_{\Delta} K_2 = P\sin
\psi$ and
$P = \sqrt{(P^1)^2 + (P^2)^2} = \sqrt{v_{F}^2 K_1^2 + v_\Delta^2 K_2^2}$.
(We denoted the components of ${\bf P}$ as $P^1, P^2$ to
make them different from the node label $P_{j}$ used in what follows.)
Finally, we can approximate the difference $n_{F}(E_{+})- n_{F}(E_{-})$
as
\begin{equation}\label{Fermi.difference}
n_{F}(E_{+})- n_{F}(E_{-}) \approx \frac{d n_{F}(E)}{d E}
{\bf v}_{g} {\bf K} =  \frac{d n_{F}(E)}{d E} P \cos (\delta - \psi).
\end{equation}
Recall also that in the nodal approximation the integral over
the original Brillouin zone is replaced by the integration over 4 nodal
sub-zones as
\begin{equation}\label{nodal.approximation}
\int \frac{d^2 k}{(2 \pi)^2} \to \sum_{j =1}^{4}\int \frac{d k_1 d k_2}{(2
\pi)^2} \to \sum_{j =1}^{4}\int \frac{d^2 p}{(2 \pi)^2 v_{F} v_\Delta} =
\sum_{j =1}^{4}\int_{0}^{p_{max}} \frac{p d p}{2 \pi v_{F} v_\Delta}
\int_{0}^{2 \pi} \frac{d \varphi}{2 \pi}\,, \quad p_{max} = \sqrt{\pi v_{F}
v_{\Delta}}/a\,.
\end{equation}
It is necessary to underline that the nodal approximation is designed for
the low temperature regime $T \ll \Delta_{0}$. Thus all polarizations derived in
Appendix~\ref{sec:C} are applicable for  $T \ll \Delta_{0}$.

\section{Calculation of $\Pi_{ij}$ and equation for $T_c$}
\label{sec:C}

After the substitution Eqs.~(\ref{difference}), (\ref{Fermi.difference})
in Eq.~(\ref{Pi_{33}}) and integration over $p$ (see \cite{we.LD}) $\Pi_{33}$
is expressed via the integral
\begin{equation}
I = \int_{0}^{2 \pi} \frac{d \delta}{2 \pi}
\frac{\cos^2 \delta \cos^2 (\delta - \psi)}{ \cos^2 (\delta - \psi) - b^2}
\end{equation}
with $b = |\Omega|/P_{j}$. (Note that we omitted the sub-zone index $j$
in $\psi$ and $b$.)
This integral can be calculated using the table
integral (3.682) from \cite{Gradshteyn} for $b^2 >1$ giving
\begin{equation}
I = \frac{1}{2} + b^2 \cos 2 \psi -  \frac{|b|^3}{\sqrt{b^2 -1}} \cos^2 \psi
 + |b|\sqrt{b^2 -1} \sin^2 \psi \,.
\end{equation}
Then using the analytical continuation in the region $b^2 < 1$,
we finally arrive at the result
\begin{equation}
\label{Pi_{33}.final}
\begin{split}
\Pi_{33} (\Omega, {\bf K}) = & - \kappa(T) -
\sum_{j=1}^{4}
\frac{\ln 2}{\pi} \frac{T}{v_F v_{\Delta}}
\left\{ \frac{1}{2}  + \frac{\Omega^2}{P_j^2 } \cos 2 \psi_j -
 \frac{|\Omega|}{P_j}
\frac{1}{\sqrt{ \frac{\Omega^2}{P_j^2} - 1 }}
\left[ \frac{\Omega^2}{P^2_j}
\cos^2 \psi_j  -
\left( \frac{\Omega^2}{P_j^2} - 1 \right)
\sin^2 \psi_j \right]
\Theta \left( \frac{|\Omega|}{P_j} - 1 \right)  \right. \\
 +  & \left. i \frac{\Omega}{P_j }
\frac{1}{ \sqrt{1- \frac{\Omega^2}{P_j^2} }}
\left[ \frac{\Omega^2}{P_j^2} \cos^2 \psi_j +
\left( 1 - \frac{\Omega^2}{P_j^2} \right) \sin^2 \psi_j \right]
\Theta \left(1 - \frac{|\Omega|}{P_j} \right) \right\} ,
\end{split}
\end{equation}
where $\kappa(T)$ denotes the term which originates from the first
term in the braces of Eq.~(\ref{Pi_{33}}) which cannot be
accurately calculated using the nodal approximation \cite{we.LD}.
It is easy to show, however, that $\kappa = m/\pi$ at $T =0$ for
the continuum $s$-wave pairing model. This can be used to write
$\Pi_{33}$ in the dimensionless form:
\begin{equation}
\label{pi_{33}.definition}
\Pi_{33}(\Omega, {\bf K}) =   \frac{m}{\pi}
{\tilde \Pi}_{33} (\Omega, {\bf K}),
\end{equation}
where
\begin{equation}
\label{pi_{33}}
{\tilde \Pi}_{33} (\Omega, {\bf K})=
- {\tilde \kappa} -
\sum_{j=1}^{4} \frac{\ln 2}{2} \frac{T}{\epsilon_F} \alpha_{D}
\left\{ \cdots \right\},
\end{equation}
where $\cdots$ are the terms in the braces in Eq.~(\ref{Pi_{33}.final}),
$\epsilon_{F} = m v_{F}^2 /2$ is the Fermi energy,
$\alpha_{D} = v_{F}/v_{\Delta}$ (see Sec.~\ref{sec:model})
and we assume that ${\tilde \kappa} \sim 1$.
One can also check that
the imaginary part of $\Pi_{33}$ coincides with the expression calculated
in \cite{we.LD} integrating $\delta(P \cos(\delta - \psi) - \Omega)$.

In the same way we arrive at the expression for
the current-current polarization function Eq.~(\ref{Pi_{00}}):
\begin{equation}
\Pi_{00}^{\alpha \beta}(\Omega, {\bf K}) K_{\alpha} K_{\beta}/K^2 =
-  \sum_{j =1}^{4} \frac{\ln 2}{\pi} \frac{T}{v_{F} v_{\Delta}}
\frac{P_{j}^{2} \cos^2 \psi_{j}}{K^2} \left[ 1 -
\frac{|\Omega|}{P_{j}}  \frac{1}{\sqrt{\frac{\Omega^2}{P_{j}^{2}} -1}}
\Theta\left(\frac{|\Omega|}{P_{j}} - 1\right)
+ i \frac{\Omega}{P_{j}} \frac{1}{\sqrt{1 - \frac{\Omega^2}{P_{j}^{2}}}}
\Theta\left(1 - \frac{|\Omega|}{P_{j}}\right).
\right]
\end{equation}
The zero order superfluid stiffness $\Lambda_{0}^{\alpha \beta} =
\delta_{\alpha \beta}n/m$ for the continuum ($m_{\alpha \beta}
(\mathbf{k}) = m \delta_{\alpha \beta}$) system at $T =0$, where
$n$ is the total carrier density which, of course, coincides with
the density of the neutralizing background. Using the expression
$\epsilon_{F} = \pi n/m$ which is, strictly speaking, valid only
for the 2D systems with the quadratic dispersion law, we can also
rewrite the superfluid stiffness $\Lambda_{\alpha \beta}$ in the
dimensionless form
\begin{equation}
\label{pi_{00}}
\Lambda^{\alpha \beta}(\Omega, \mathbf{K})
\frac{K_{\alpha} K_{\beta}}{K^2} \equiv
\frac{\epsilon_{F}}{\pi} {\tilde \Lambda}
= \frac{\epsilon_{F}}{\pi} \left[
{\tilde \Lambda_{0}} - \sum_{j=1}^{4} \ln 2 \frac{T}{\epsilon_{F}} \alpha_{D}
\frac{P_{j}^{2}}{(v_{F} K)^{2}} \cos^2 \psi_{j} \cdots
\right]\,.
\end{equation}
As was already mentioned after Eq.~(\ref{Pi_{33}.final}), the
terms which contain  the averaging over the Fermi surface (see
e.g. Eqs.~(\ref{Lambda0}) and (\ref{n(k)})) cannot be accurately
calculated using the nodal approximation \cite{we.LD}. Thus, in
general  ${\tilde \Lambda_{0}}$ as well as ${\tilde \kappa}$
should be considered as a free parameter of the model. In
particular, decreasing  the value of ${\tilde \Lambda_{0}}$ it is
possible to describe a lowering of the zero temperature superfluid
stiffness in HTSC. Nevertheless, for the numerical computations we
will assume that ${\tilde \Lambda_{0}} \sim 1$. It is easy to
obtain (see e.g. \cite{Lee,we.LD}) that the static, zero momentum
bare superfluid stiffness
\begin{equation}
\label{stiffness.Tc}
\Lambda = \frac{\epsilon_{F}}{\pi}
\left( {\tilde \Lambda_{0}}  - 2 \ln 2 \alpha_{D}\frac{T}{\epsilon_{F}} \right)
\end{equation}
and that the velocity of the BA mode
\begin{equation}
\label{v}
v = \sqrt{\frac{\Lambda}{\Pi_{33}(0, \mathbf{K})}} =
\frac{v_{F}}{\sqrt{2}}  \sqrt{\frac{{\tilde \Lambda_{0}}-
 2 \ln 2 \alpha_{D}T /\epsilon_{F}}{{\tilde \kappa}}}\,.
\end{equation}
so that for ${\tilde \Lambda_{0}} = {\tilde \kappa} = 1$ the BA mode velocity
$v(T =0) = v_{F}/\sqrt{2}$.
Using Eq.~(\ref{stiffness.Tc}) one can estimate
the temperature of  Berezinskii-Kosterlitz-Thouless transition
from the equation $T_c = \pi / 2 \Lambda(T_c)$, which gives
\begin{equation}
\label{Tc}
T_c = \frac{\epsilon_F {\tilde \Lambda_0}}{2 (1 + \ln 2 \alpha_{D})}.
\end{equation}
We will use this definition of $T_c$ to express the temperature
$T$ in the units of $T_c$ and $\epsilon_{F}$.

Finally, we obtain that the expression for Eq.~(\ref{Pi_{03}})
\begin{equation}
\label{pi_{03}}
\Pi_{03}^{\alpha}(\Omega, {\bf K}) K_{\alpha}/K = - \sum_{j =1}^{4}
\frac{T}{\pi v_{F}} \left\{ \alpha_{D} \ln 2 \frac{\Omega}{v_{F} K}
\cos \psi_{j} \left[ \cos \psi_{j} - \frac{|\Omega|}{P_{j}}
\frac{\cos \psi_{j}}{\sqrt{ \frac{\Omega^2}{P_{j}^{2}} -1}}
\Theta\left(\frac{|\Omega|}{P_{j}} -1 \right)
+ i \frac{\Omega}{P_{j}}
\frac{\cos \psi_{j}}{\sqrt{ 1 - \frac{\Omega^2}{P_{j}^{2}}}}
\Theta\left(1 - \frac{|\Omega|}{P_{j}}\right)
\right]\right\}\,,
\end{equation}
where we put inside the braces the dimensionless part,
${\tilde \Pi_{03}}^{\alpha} K_{\alpha}/K$.

\section{Equations for $v_{CG}$, structure factor and transformation
to the global coordinate system}
\label{sec:D}

Substituting Eqs.~(\ref{pi_{33}.definition}), (\ref{pi_{00}}) and (\ref{pi_{03}})
in Eq.~(\ref{CG.equation}), we obtain the equation
for CG mode in  the dimensionless form which is convenient for numerical
investigation
\begin{equation}\label{CG.dim}
{\tilde \Lambda} = \frac{T^2}{2\epsilon_{F}^2}
\frac{({\tilde \Pi_{03}}^{\alpha} K_{\alpha}/K)^2}
{{\tilde \Pi_{33}}}\,.
\end{equation}

The whole expression (\ref{theta.cor.new}) for $M_{\theta}^{-1}$
can also be written as
\begin{equation}
\label{M.theta.num}
M_{\theta}^{-1}(\Omega, \mathbf{K}) = \frac{m}{\pi} v_{F}^2 K^2
{\tilde M}_{\theta}^{-1}(\Omega, \mathbf{K}),
\end{equation}
where
\begin{equation}
\label{M.tilde}
{\tilde M}_{\theta}^{-1}(\Omega, \mathbf{K}) =  \frac{\Omega^2}{v_F^2 K^2}
\mathbf{\hat \Pi}_{33}(\Omega, \mathbf{K}) - \frac{\Omega}{v_F K}
\mathbf{\hat \Pi}_{03}^{\alpha}(\Omega, \mathbf{K}) \frac{K_{\alpha}}{K}
+ \frac{1}{2}\mathbf{\hat \Lambda}^{\alpha \beta} \frac{K_{\alpha} K_{\beta}}{K^2},
\end{equation}
where the dimensionless polarization functions  $\mathbf{\hat \Pi}_{33}$,
$\mathbf{\hat \Pi}_{03}^{\alpha}$ and $\mathbf{\hat \Lambda}^{\alpha \beta}$
were made from the full polarization functions $\mathbf{\bar \Pi}_{33}$,
$\mathbf{\bar \Pi}_{03}^{\alpha}$, $\mathbf{\bar \Lambda}^{\alpha \beta}$
(see Eqs.~(\ref{phase.plasmon}) and (\ref{theta.cor.new})) in the same
way as the polarizations Eqs.~(\ref{pi_{33}.definition}), (\ref{pi_{00}}) and
(\ref{pi_{03}}). The only difference is that these full polarizations include
the Coulomb potential $V_{c}(\mathbf{K})$ (for simplicity we take the
3D potential) which for our purposes is convenient to rewrite as
\begin{equation}
\label{Coulomb}
V_{c}^{-1} (\mathbf{K}) = \frac{m}{2 \pi} \frac{v_F^2 K^2}{\omega_p^2}\,,
\end{equation}
where $\omega_p$ is the plasma frequency defined after
Eq.~(\ref{current-density.general}).

Although the local nodal coordinate systems $(P_j, \psi_{j})$ are
very convenient for calculating of the polarization functions
Eqs.~(\ref{pi_{33}}), (\ref{pi_{00}}) and (\ref{pi_{03}}), the
final expressions for them and, for example,
Eq.~(\ref{M.theta.num}) have to be calculated in the global or
laboratory coordinate system $(K, \phi)$. It is convenient to
measure the angle $\phi$ from the vector $\hat{k}_x$, so that
$\phi =0$ corresponds to the corner of the Fermi surface (see e.g.
Fig.~1 in \cite{we.LD}) and the first node is at $\phi = \pi/4$.
Thus the transformations from the global coordinate system into
the local system related to the $j$-th node are
\begin{equation}\label{transform}
\begin{split}
& P_j = K \sqrt{v_F^2 \cos^2 \left(\phi -\frac{\pi}{4}+
\frac{\pi}{2}(j-1)\right) +
v_{\Delta}^2 \sin^2 \left(\phi - \frac{\pi}{4}+ \frac{\pi}{2}(j-1) \right)}\,, \\
& \cos \psi_j = \frac{v_F K}{P_j} \cos \left(\phi - \frac{\pi}{4}+
\frac{\pi}{2}(j-1)\right)\,, \qquad
\sin \psi_j = \frac{v_{\Delta} K}{P_j} \sin
\left(\phi - \frac{\pi}{4} + \frac{\pi}{2}(j-1)\right)\,,
\qquad j =1, \ldots, 4.
\end{split}
\end{equation}

%\newpage

\begin{thebibliography}{99}
\bibitem{foot1} ``Low energy'' means that the energy of the phase
distortions is smaller than than the enery gain, $E_{\rm cond}$
due to the gap opening for neutral or
$\Omega^{2} \lesssim E_{\rm cond} V_{c}(\mathbf{K})$ for
charged system \cite{Randeria.action}.
\bibitem{foot3} For completness we reintroduced for a moment the Planck's
constant $\hbar$. Note also that in the conventions that we used
the charge of electron is $e = - |e|.$
\bibitem{foot5} In principle, a more general equation (\ref{phase.modes})
without taking its real part has to be studied to extract the information
about the damping of the CG mode, but this significantly increases the difficulty
of the numerical work. From this point of view a direct computation of the
structure factor which does not demand solving any equation is more
straightforward.
\end{thebibliography}
\end{document}